\documentclass[aps,prb,preprint,superscriptaddress,showkeys,longbibliography]{revtex4-1}
\usepackage{natbib,graphicx,dcolumn,color,lipsum} 
\usepackage{subfigure,amsmath,amssymb,verbatim,moreverb,bm,framed}
\usepackage{gensymb,ulem} 
\usepackage{xcolor}

\usepackage[pdftex,
]{hyperref}

\def\be{\begin{equation}}
	\def\ee{\end{equation}}
\def\ber{\begin{eqnarray}}
	\def\eer{\end{eqnarray}}
\newcommand{\Tdos}{{\cal{D}}}	
\newcommand{\ie}{{\it i.e.}} 	
\newcommand{\eg}{{\it e.g.}} 	

\begin{document}

\title{2D-Dirac surface states and bulk gap probed via quantum capacitance in a 3D topological insulator}

\author{Jimin Wang}
\affiliation{Institute of Experimental and Applied Physics, University of Regensburg, 93040 Regensburg, Germany}
\author{Cosimo Gorini}
\affiliation{Institute of Theoretical Physics, University of Regensburg, 93040 Regensburg, Germany}
\author{Klaus Richter}
\affiliation{Institute of Theoretical Physics, University of Regensburg, 93040 Regensburg, Germany}
\author{Zhiwei Wang}
\affiliation{Physics Institute II, University of Cologne, Z\"ulpicher Str. 77, 50937 K\"oln, Germany}
\affiliation{Key Laboratory of Advanced Optoelectronic Quantum Architecture and Measurement, Ministry of Education, School of Physics, Beijing Institute of Technology, Beijing, 100081, China}
\author{Yoichi Ando}
\affiliation{Physics Institute II, University of Cologne, Z\"ulpicher Str. 77, 50937 K\"oln, Germany}
\author{Dieter Weiss}
\affiliation{Institute of Experimental and Applied Physics, University of Regensburg, 93040 Regensburg, Germany}

\date{\today}

\begin{abstract}
  BiSbTeSe$_2$ is a 3D topological insulator (3D-TI) with Dirac type surface states and low bulk carrier density, as donors and acceptors compensate each other. Dominating low temperature surface transport in this material is heralded by Shubnikov-de Haas oscillations and the quantum Hall effect. Here, we experimentally probe and model the electronic density of states (DOS) in thin layers of BiSbTeSe$_2$ by capacitance experiments both without and in quantizing magnetic fields. By probing the lowest Landau levels, we show that a large fraction of the electrons filled via field effect into the system ends up in (localized) bulk states and appears as a background DOS. The surprisingly strong temperature dependence of such background DOS can be traced back to Coulomb interactions. Our results point at the coexistence and intimate coupling of Dirac surface states with a bulk many-body phase (a Coulomb glass) in 3D-TIs.
\end{abstract}
\maketitle

\label{Intro}

An ideal three-dimensional (3D) topological insulator (TI) is a band insulator, characterized by a gap in the single-particle energy spectrum, and symmetry protected, conducting surface states \cite{HasanRMP2010,AndoJPSJ2013} (and references therein). Experimentally available TIs like Bi$_2$Se$_3$ or Bi$_2$Te$_3$ are, however, far from ideal as they feature, due to intrinsic defects, a relatively high electron or hole density larger than 10$ ^{18} $ cm$ ^{-3} $ (see \cite{AndoPRB2016_OpticalConduc.Puddles} and references therein). By combining \textit{p}-type and \textit{n}-type TI materials, \ie, by compensation, the bulk concentration can be suppressed \cite{AndoJPSJ2013,Ren2011}. This comes at the price of large potential fluctuations at low temperatures as the resulting ionized donor and acceptor states are poorly screened and constitute a randomly fluctuating Coulomb potential, bending the band edges and creating electron and hole puddles \cite{SkinnerPRL2012,SkinnerJETP2013}. These were observed by, \eg, optical spectroscopy \cite{AndoPRB2016_OpticalConduc.Puddles} and scanning tunneling experiments\cite{KnispelPRB2017}.
In the absence of metallic surface states, \ie, in compensated conventional semiconductors, variable range hopping governs low-temperature transport ($T<$ 100\,K) \cite{EfrosShklovskii1984,SkinnerPRL2012}. Recently, Skinner \textit{et al.} have shown that the electronic density of states (DOS) in the bulk is nearly constant under these circumstances and features a Coulomb gap at the Fermi level \cite{SkinnerPRL2012,SkinnerJETP2013}. In 3D-TIs, in addition, Dirac surface states, which form a two-dimensional (2D) electron (hole) system, encase the bulk and constitute the dominating transport channel at low temperatures. 

The nature of the surface and bulk phases is antithetical. The helical surface metal is an example of Berry Fermi liquid \cite{chen2017}, whose constituents are resilient to Anderson localization \cite{HasanRMP2010} and well described within a single (quasi-)particle picture.  Bulk electrons on the other hand are topologically trivial, and organize into a many-body disordered and localized phase. 
Indeed, such a phase shows characteristics \cite{SkinnerJETP2013,meroz2014} typically ascribed to a Coulomb glass, an exotic insulating state known and studied for decades, yet far from being fully understood \cite{davies1984,vignale1987,ovadyahu2013}.  The interplay between these two different phases is largely unexplored ground, one reason being the difficulty in engineering a system where both coexist.  A largely-compensated 3D-TI like BiSbTeSe$_2$ seems however ideally suited for this purpose, appearing as an intrinsic two-phase hybrid system. Moreover, understanding the surface-bulk interplay is not only interesting for fundamental reasons, but also necessary if BiSbTeSe$_2$ -- and more generally (fully-)compensated 3D-TIs -- is used as a device platform to realize, for example, topological superconductivity, Majorana zero modes \cite{FuKane2008,FuKane2009}, and topological magnetoelectric effects \cite{EssinPRL2009}. With this goal in mind, we explore the surface-bulk interplay by probing the DOS of the Dirac surface states. 

The method we used is capacitance spectroscopy, which provides complementary information, compared with common transport measurements. The total capacitance $ C $, measured between a metallic top gate and the Dirac surface states, depends on the geometric capacitance, $C_0=\epsilon \epsilon_{0}A/d $, and the quantum capacitance  $ Ae^{2} D(\mu) $, 
\be
\label{eq_QC1}
C^{-1}= C_0^{-1}+ \left[Ae^{2}D(\mu)\right]^{-1}.
\ee
Here $\epsilon, d$, $A$ are, respectively, relative dielectric constant, and thickness of the insulator, as well as capacitor area, $ \epsilon_{0}$ the vacuum dielectric constant, and $D(\mu)$ the DOS at the Fermi level (chemical potential) $\mu$. The quantum capacitance, connected in series to $C_0$, reflects the energy spectrum of 2D electron systems \cite{TPSmith1985,VMosser1986,PonomarenkoPRL2010_GrapheneQC}, and probes preferentially the top surface DOS in 3D-TIs \cite{KozlovPRL2016_QC}. At higher temperatures, $D(\mu)$ has to be replaced by the thermodynamic density of states (TDOS) at $\mu$, $\Tdos(\mu) = dn/d\mu$, with $n$ the carrier density. 
While gating of 3D-TIs and tuning of the carrier densities of top and bottom surfaces \cite{AshooriPRL2014_ElectroStaticCouplingTI,Xiong2013, ChongPRL2019}, and even magnetocapacitance \cite{ChongPRL2019, ChongACSNano2020} has been explored in the past, the analysis of quantum capacitance and the DOS in a compensated TI like BiSbTeSe$_2$ remained uncharted. Our measurements show that, while Dirac surface states dominate low-$T$ transport as expected, the bulk provides a background, capable of absorbing a large amount of charge carriers.
These missing charges are very common in 3D-TI transport experiments, yet to the best of our knowledge unexplained \cite{XuNC2016,ZhangNC2017_CoDecoratedBSTS,ChongNanoLett2018_CGT+BSTS,YoshimiNC2015_BST_QHE}. 
Furthermore, our in-depth analysis of the quantum capacitance data shows that the background is not a rigid object, but reorganizes itself depending on the gate voltage (and temperature) value.  This is a signature of a strongly interacting many-body phase, whose {\it effective} single-particle DOS reshapes itself adapting to varying conditions
--  \eg\,when charges are added or removed.  The reshaping is typically slow, with a reaction/relaxation time which can be orders of magnitude slower than that of the surface Dirac metal.
This provides evidence that BiSbTeSe$_2$ is not an ideal 3D-TI, but rather a hybrid two-phase system, and suggests that similar behavior should be expected for other compensated 3D-TI materials.

\textit{Transport measurements} -- The details of sample fabrication and measurements can be found in Supporting Information. Fig. 1(a) and 1(b) display the layer sequence and an optical micrograph of one of the devices, respectively. Temperature dependent measurements show (see Fig.~S2 in the Supporting Information) that at 1.5\,K transport is entirely dominated by the surface with negligible contribution from the bulk. The carrier density and $\mu$ of top and bottom surfaces can be adjusted by top and bottom gate voltages, $V_{\text{tg}}, V_{\text{bg}}$, respectively. This is shown for the Hall resistivity at 14 T in Fig.~1(c). The device displays well developed quantum Hall plateaus at total filling factor $\nu=$ -1, 0 and 1 (Note that for each Dirac surface state, Landau levels are fully filled at half integer $\nu$ and total filling factor 1, corresponding to $\nu\,=\,1/2$ on top and bottom surfaces \cite{XuNC2016}.). The plateaus are well separated from each other and marked by dashed purple lines. As these lines run nearly parallel to the $V_{\text{tg}}$- and $V_{\text{bg}}$-axes, respectively, we conclude that the carrier density on top and bottom can be tuned nearly independently.

\begin{figure} 
	\includegraphics[width=\linewidth]{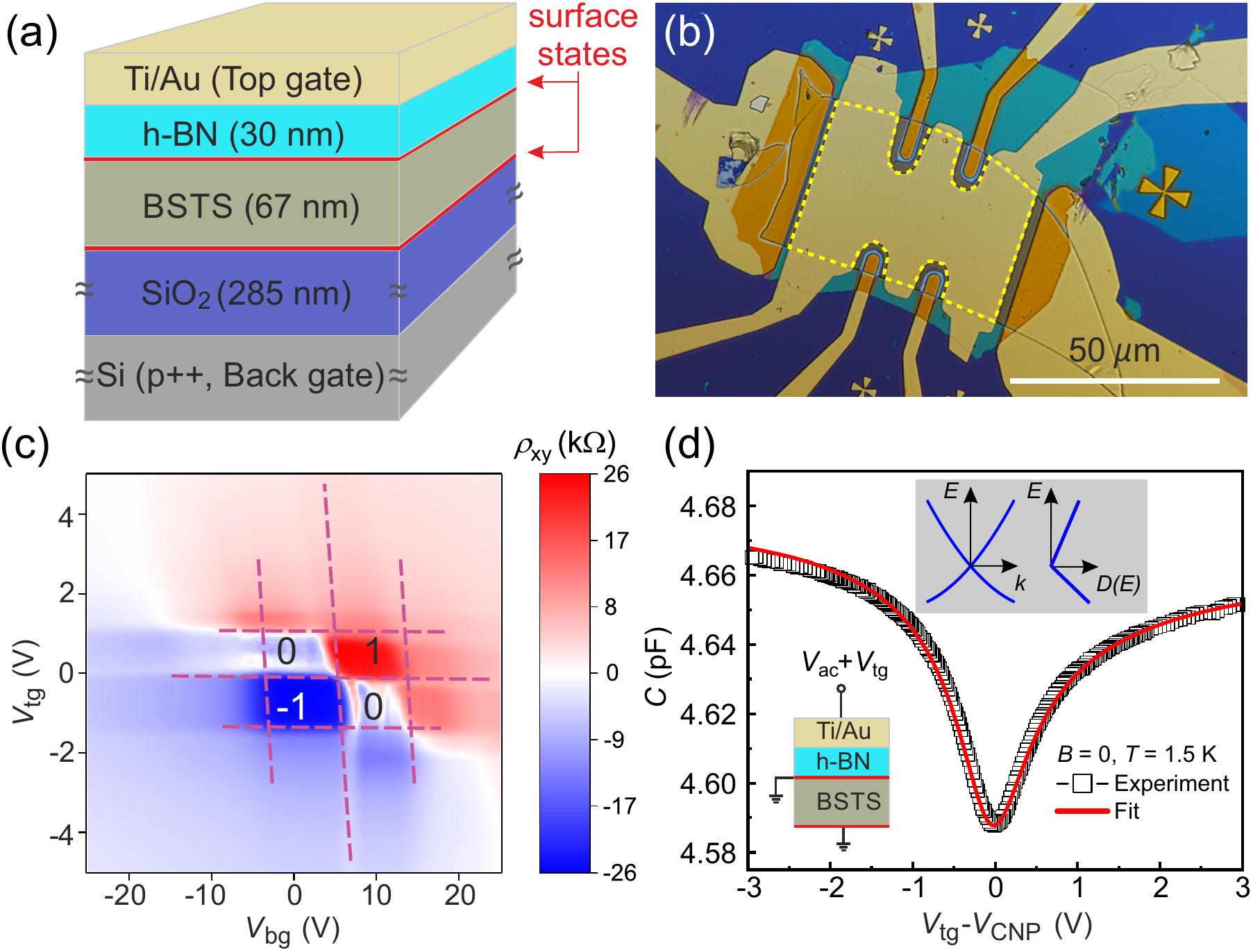}
	\caption{
		(a) Design of the layer sequence. Red lines sketch the topological surface states. (b) Optical micrograph of the device. The dashed yellow line marks the capacitor area of about $1.8\times10^3\,\mu \text m^2$. (c) $\rho_{\text{xy}}$ as function of $V_{\text{tg}} $ and $V_{\text{bg}}$, respectively, at $T=1.5$\,K and $B=14$\,T. The almost horizontal and vertical dashed purple lines separate the region of well developed QHE with total filling factors of -1, 0, and 1 (unit $h/e^2$), from regions of  higher filling factors. (d)  $C(V)$ at $T=1.5$\,K and $B=0$\,T. Unlike graphene, the two branches are asymmetric with respect to CNP due to a non-linear $E-k$ relation. The pronounced minimum reflects the bulk gap. For better comparison, the CNP of all measurements is shifted to zero via $V_{\text{CNP}}$.  The red line is a fit using a Gaussian broadening of the Fermi level with $\sigma = 29.4$\,meV (see text). To compare with experiment we added a parasitic capacitance (mainly coming from wiring and bonding in the immediate vicinity of the device) of $\sim$3.07\,pF. The lower left inset illustrates the measurement configuration, the upper inset shows the energy dispersion of the surface states in the bulk gap (left) and the corresponding DOS (right).
	}
	\label{fig_device}
\end{figure}

\textit{Capacitance measurements} -- Figure 1(d) shows the measured capacitance $C$, which is directly connected to the DOS, see Eq.~\eqref{eq_QC1} (measurement details are in the Supporting Information). The measured trace with a minimum at the Dirac or charge neutrality point (CNP) resembles the quantum capacitance measured for graphene, apart from a pronounced electron-hole asymmetry \cite{PonomarenkoPRL2010_GrapheneQC,YuPNAS2013_GrapheneQC,
XiaNatNano2009_GrapheneQC} due to a parabolic contribution to the linear $E(k)$ dispersion \cite{AndoPRL2011_Ek}.  Explicitly, the latter reads $E=\pm\hslash v_{\rm F} k +  \dfrac{\hslash ^{2} k^{2}}{2m^{*}}$, with $\hbar$ the reduced Planck constant, $v_{\rm F}$ the Fermi velocity at the Dirac point, and $m^*$ the effective mass.  It is sketched in the upper inset of Fig.~1(d), together with the electron-hole asymmetric, nearly $E$-linear DOS, given by $D(E)=\left | \frac{m^{*} (\Omega-m^{*} v_{\rm F})}{2\pi \hslash^2 \Omega} \right | $. Here we used that $k=\sqrt{4\pi n}$ and  $  \Omega=\sqrt{(m^{*}v_{\rm F})^2+2Em^*} $. 

While in a perfect system $D(E)$ vanishes at the CNP, disorder smears the singularity, as in case of graphene \cite{PonomarenkoPRL2010_GrapheneQC}.
We model the potential fluctuations by a Gaussian distribution of energies with width $\sigma$, resulting in an average DOS $\langle D(\mu) \rangle  =\int_{-\infty}^{\infty}{D(E)\frac{1}{\sqrt{2 \pi }\sigma} \exp\big[\frac{-(E-\mu)^2}{2\sigma^2}\big]}dE$. To convert energies into voltages we use $n=C_0 (V_{\text{tg}}-V_0)/Ae$, with $e$ the elementary charge and $V_0$ describing $n$ at zero voltage.
By fitting $\langle D(\mu) \rangle$ to the data in Fig. 1(d) we extract $\sigma = 29.4$ meV, $v_{\rm F} =3.2\cdot 10^{5}$\,m/s and $m^*=0.47 m_0$ ($m_0=$ free electron mass). The broadening $\sigma$, which is in reasonable agreement with theory\cite{SkinnerPRB2013}, is only important in the immediate vicinity of the CNP but hardly affects the values of $v_{\rm F}$ and $m^*$. The obtained $ v_{\rm F} $ and $ m^{*} $ values agree well with ARPES data \cite{AndoNC2012_ARPES_BSTS,AshooriPRL2014_ElectroStaticCouplingTI} and values extracted from Shubnikov-de Haas oscillations \cite{AndoPRL2011_Ek}. 

\textit{$B$-field dependence of capacitance measurements} -- Our key result arises when we crank up the magnetic field and measure signatures of the Landau level (LL) spectrum, shown in Fig.~2(a). At the 0-th LL level position a local maximum emerges with increasing $ B $-field, flanked by minima at each side. The two minima, highlighted by arrows, correspond to the Landau gaps between LLs 0, and $\pm 1$ (see Fig.~2(a)). Due to the large broadening, higher LLs do not get resolved.
Lowering $T$ down to 50\,mK does not resolve more structure, indicating that disorder broadening is the limiting factor. By sweeping $V_{\text{tg}}$ across the 0-th LL, \ie, from arrow position to arrow position, the carrier density changes by the LL degeneracy $\Delta n = eB/h $. In contrast, the change of carrier density $\Delta n$, calculated via capacitance, $\Delta n = \frac{C_0}{Ae}\Delta V_{\text{tg}}$, is by a factor 1.4 higher. Hence, we must assume that a large fraction of the carriers, induced by field effect, ends up in the bulk and is localized at low $T$.

To compare with these experiments we calculate $C(V_{\text{tg}})$
using Gaussian-broadened LLs, 
$D_{\rm LL}(E)=\frac{1}{\sqrt{2 \pi }\Gamma} \frac{eB}{h} \sum \exp\big[\frac{-(E-E_{\rm n})^2}{2\Gamma^2}\big]$, with broadening $\Gamma$. The LL spectrum dispersion reads \cite{AndoPRB2011_Ei} 
\be
\label{LLspectrum}
E_{\rm n} = \lvert n \rvert \frac{\hbar eB}{m^*} + {\rm sgn}(n) 
\sqrt{\left(\frac{\hbar eB}{2m^*}\right)^2+2eB\hbar v_{\rm F}^{2} \lvert n \rvert },
\ee
with $n=0,\,\pm1,\,\pm2,\,\dots$, and the tiny Zeeman splitting was neglected. 
Using the above DOS is insufficient to describe the data - the calculated distance $\Delta V_{\text{tg}}$ between adjacent Landau gaps is too small and does not match the minima positions observed in experiment (marked by arrows in Fig. 2(a) for the 14 T trace, see Fig. S4 in the Supporting Information as an example).
$\Delta V_{\text{tg}}$ is the voltage needed to fully fill the 0-th LL of the surface states. Since $\Delta V_{\text{tg}}$ in experiment is larger than that in calculation (relies on the filling rate $ dn/dV_{\text{tg}} \approx C_0/Ae $), it means that a fraction of the field-induced electrons does not go to the surface states but eventually into the bulk. Thus a higher voltage (higher $\delta n$) is needed to fill the zeroth LL. In contrast, we obtain almost perfect agreement -- see Fig.~\ref{fig_B_dependent_QC}(b) -- if we introduce an
energy-independent background DOS $ D_{\rm b} $ which models these bulk states.
The calculated TDOS we compare with experiment thus reads
\begin{equation}
\label{TDoS}
\Tdos(\mu)=\int_{-\infty}^{\infty}{\left[D_{\rm LL}(E)+D_{\rm b}\right]\frac{\partial f}{\partial \mu}dE},
\end{equation}
with $f=f(E-\mu,T)$ the Fermi function.

\begin{figure}
	\includegraphics[width=\linewidth]{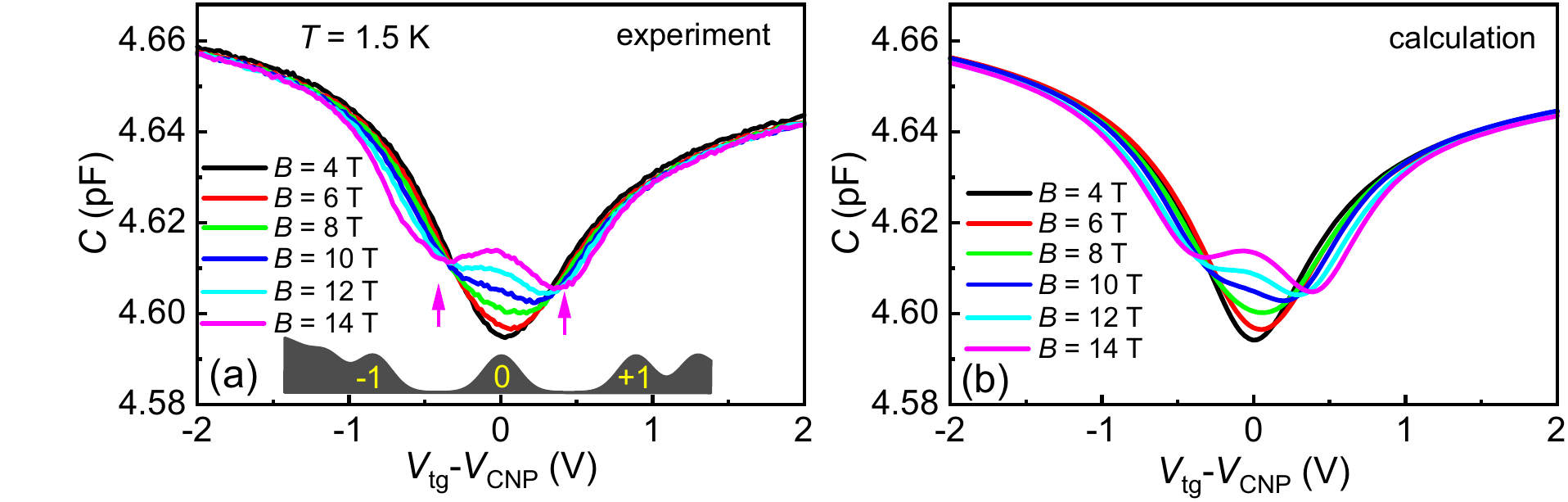}
	\caption{
		(a) $C(V_{\rm tg})$ for $B$ ranging from 4\,T to 14\,T.  The lower inset sketches the LL DOS for LLs -1, 0 and 1. Arrows mark the position of the Landau gaps for the 14\,T trace. (b) Model calculations to (a) based on Eq. \eqref{TDoS} after adding a parasitic capacitance of $\sim 3.06$\,pF. Parameters of the fit at 1.5\,K:  $\Gamma = 14.9$ to $15.9$\,meV;  $D_{\rm b} = 2.4 \cdot 10^{35}\, \text{m}^{-2}\text{J}^{-1}$. Although the position of the 0-th LL changes slightly with $B$, we use the same zero-field value $V_{\rm CNP}$ for all curves.
	}
	\label{fig_B_dependent_QC}
\end{figure}

As shown in Fig.~2(b), the constant background $D_{\rm b}$ leads to excellent agreement with experiment. Although the bulk DOS is hardly directly accessible by the quantum capacitance itself (\ie, by its value),
we probe it indirectly via the missing charge carriers given by the Landau gap positions.
This missing charge carrier issue holds also for the quantum Hall trace in Fig.~1(c), 
where $ \sim $\,30\% of the induced electrons are missing.
Indeed, it also appears in several other publications 
with missing electron fractions ranging from 30\% (as here) to 75\% 
(see \cite{XuNC2016,ZhangNC2017_CoDecoratedBSTS,ChongNanoLett2018_CGT+BSTS,YoshimiNC2015_BST_QHE}).

The bottom line is: The change of surface carrier density extracted from the Landau gap positions is smaller than the one "loaded" into the system within the same voltage interval. 
Further, the filling rate $ dn/dV_{\text{tg}} $ determined by the classical Hall effect at 1.5\,K is consistent with the one found for the surface states (see the Supporting Information). Thus, the charge carriers loaded at low $T$ into the bulk are localized and do not contribute to transport. This is consistent with transport experiments \cite{XuNatPhys2014, XuNC2016}, and also in line with what is expected in compensated semiconductors \cite{EfrosShklovskii1984}, as was recently highlighted in Ref.~\cite{SkinnerPRL2012}. There, bulk transport of compensated TI was considered,
where local puddles of \textit{n}- and \textit{p}-regions form.  
In this regime, low-$ T $ transport is governed by variable range hopping, the DOS is, apart from the Coulomb gap, essentially constant for perfect compensations  but changes its form strongly if the chemical potential shifts. \cite{EfrosShklovskii1984,SkinnerPRL2012,SkinnerJETP2013}

Using a constant background affects somewhat the values extracted above from $C(V_{\text{tg}},B=0)$.
Thus, we fitted the trace in Fig.~\ref{fig_device}(d) using the same $D_{\rm b} = 2.4\cdot10^{35}\ \text{m}^{-2}\text{J}^{-1}$. Now a reduced broadening $\sigma = 15$\, meV is needed, which is still in reasonable agreement with theory\cite{SkinnerPRB2013}. $C(V_{\text{tg}})$ is then best described by slightly modified values: $v_{\rm F} =2.8\cdot 10^{5}$\,m/s and $m^*=0.57 m_0$, respectively, 
still compatible with results reported elsewhere 
\cite{AndoNC2012_ARPES_BSTS,AshooriPRL2014_ElectroStaticCouplingTI,AndoPRL2011_Ek}. 

\begin{figure}
	\includegraphics[width=\linewidth]{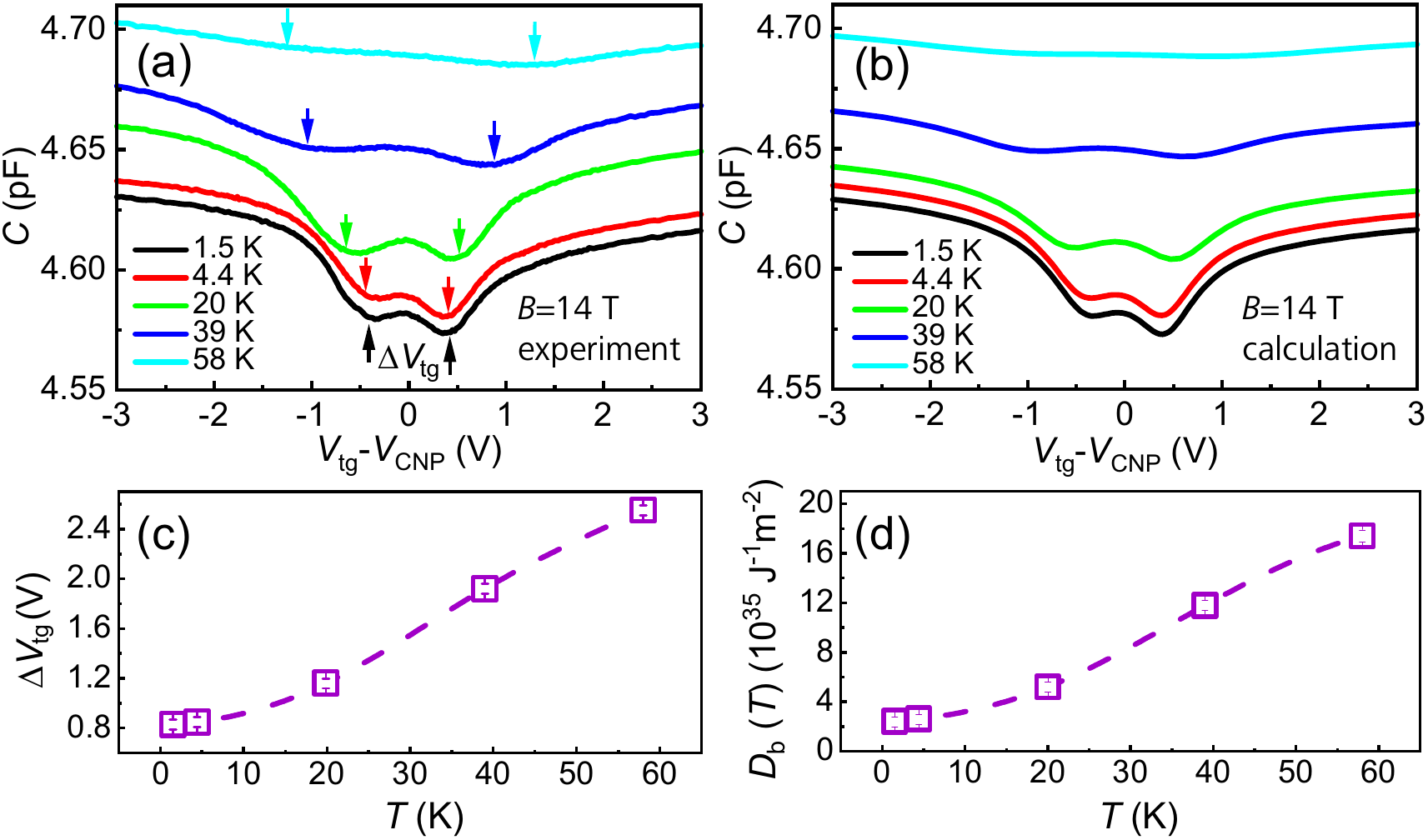}
	\caption{
		(a) $C(V_{\text{tg}}$) at $B=14$ T for various $T$s. Arrows mark the minima corresponding to Landau gaps. For increasing $T$ the voltage difference $\Delta V_{\text{tg}}$ between adjacent gaps increases. The trace at 1.5 K was shifted down by 0.032 pF for clarity. (b) Calculated $C(V_{\text{tg}}$)  using Eq.~\eqref{TDoS} with $D_{\rm b}$ values in (d), and $\Gamma = 13$ to 15.2\, meV. A nearly $T$-independent parasitic capacitance of 3.06\,$\pm$\,0.04 pF is used to best fit the data. (c) $\Delta V_{\text{tg}}$ vs. $T$. (d) Extracted  $D_{\rm b}$. The dashed lines in (c) and (d) are guides to the eye.
	}
	\label{fig_T_dependent_QC}
\end{figure}

\textit{Temperature dependence of quantum capacitance} -- The background DOS rises quickly with temperature. Corresponding $C(V_{\text{tg}})$ data for 14\,T and various $T$s 
up to 58\,K is shown in Fig.~\ref{fig_T_dependent_QC}(a). The local minima due to Landau gaps, marked by arrows, shift with increasing $T$ to larger $V_{\text{tg}}$. The corresponding $\Delta V_{\text{tg}}(T)$ is shown in Fig.~\ref{fig_T_dependent_QC}(c).  For fixed $B$ the Landau degeneracy $eB/h$ is constant and does not depend on temperature. The increasing $\Delta V_{\text{tg}}$ needed to fill the 0-th LL of the surface states thus indicates that, with increasing $T$, more carriers are lost to the bulk.
Similar behavior was found for quantum Hall data \cite{XuNC2016,XuNatPhys2014}. 
Clearly, to model the Landau gap positions correctly, a strongly $T$-dependent thermodynamic density of states (TDOS) is required.
A simple approach consists in introducing a $T$-dependent background DOS, $D_\text{b} \to D_\text{b}(T)$. 
Its values used to fit the data of Fig.~\ref{fig_T_dependent_QC}(a) 
are shown in Fig.~\ref{fig_T_dependent_QC}(d); the resulting $C(V_{\text{tg}})$ traces for different temperatures are plotted 
in Fig.~\ref{fig_T_dependent_QC}(b).  $D_\text{b}$ is nearly constant at low $T$ but rises quickly at higher temperatures, as shown in Fig.~\ref{fig_T_dependent_QC}(d).  However, a closer inspection of the possible microscopic origins 
of $D_\text{b}(T)$ reveals the central issue hiding behind our data: 
How can the TDOS near the Dirac point (at 14 T the Landau gaps are located between -30 meV and +30 meV, as estimated by Eq.~(\ref{LLspectrum})) be at the same
time practically flat, yet so strongly $T$-dependent?   
One could attempt to explain its constant value at 1.5 K by conventional trapped surface states between BSTS and hBN. They are likely responsible for the small hysteresis observed when sweeping $V_\text{tg}$, 
but the $C$-$V$ trace shift for up- and down-sweep first stays constant at low $T$, then drops with increasing temperature (see the Supporting Information). This rules out the interface states as reason of the striking TDOS increase with $T$. 
Alternatively, one could argue that the increased TDOS stems from thermal smearing of the DOS of the (effective) band edges, 
separated by a reduced gap of $\sim 60\,$meV, as determined by the measured activation energy (see the Supporting Information), instead of the full gap of 300 meV. 
Modeling this scenario by choosing the DOS at the band edges such that the average TDOS, 
when sweeping $\mu$ from one Landau gap to the other, equals the extracted constant $D_\text{b}$ cannot explain the experimental traces: The resulting TDOS is strongly energy-dependent, reflecting the sharp DOS shoulders at the band edges,
and this strong dependence would completely dominate the capacitance signal (Fig. S10 in the Supporting Information).
Thus, we are unable to find a single-particle DOS which is consistent with the experimental findings,
suggesting that a single-particle picture is simply not adequate.

\textit{Probing the many-body background} -- A way out of this apparent dead-end is provided by the strongly fluctuating potential landscape of compensated 
TIs like BiSbTeSe$_2$, sketched in Fig.~\ref{fig_puddles}(a)-(b), where Coulomb interaction dominates \cite{SkinnerPRL2012}: In a nutshell,
the background DOS emerges as an {\it effective} single particle DOS describing the ensemble of strongly interacting
electrons filling bulk impurity states \cite{EfrosShklovskii1984}.  
As such, it is actually a $\mu$- and $T$-dependent object, $D_{\rm b}\to D_{\rm b}(E,\mu,T)$, 
\begin{figure}
	\includegraphics[width=\linewidth]{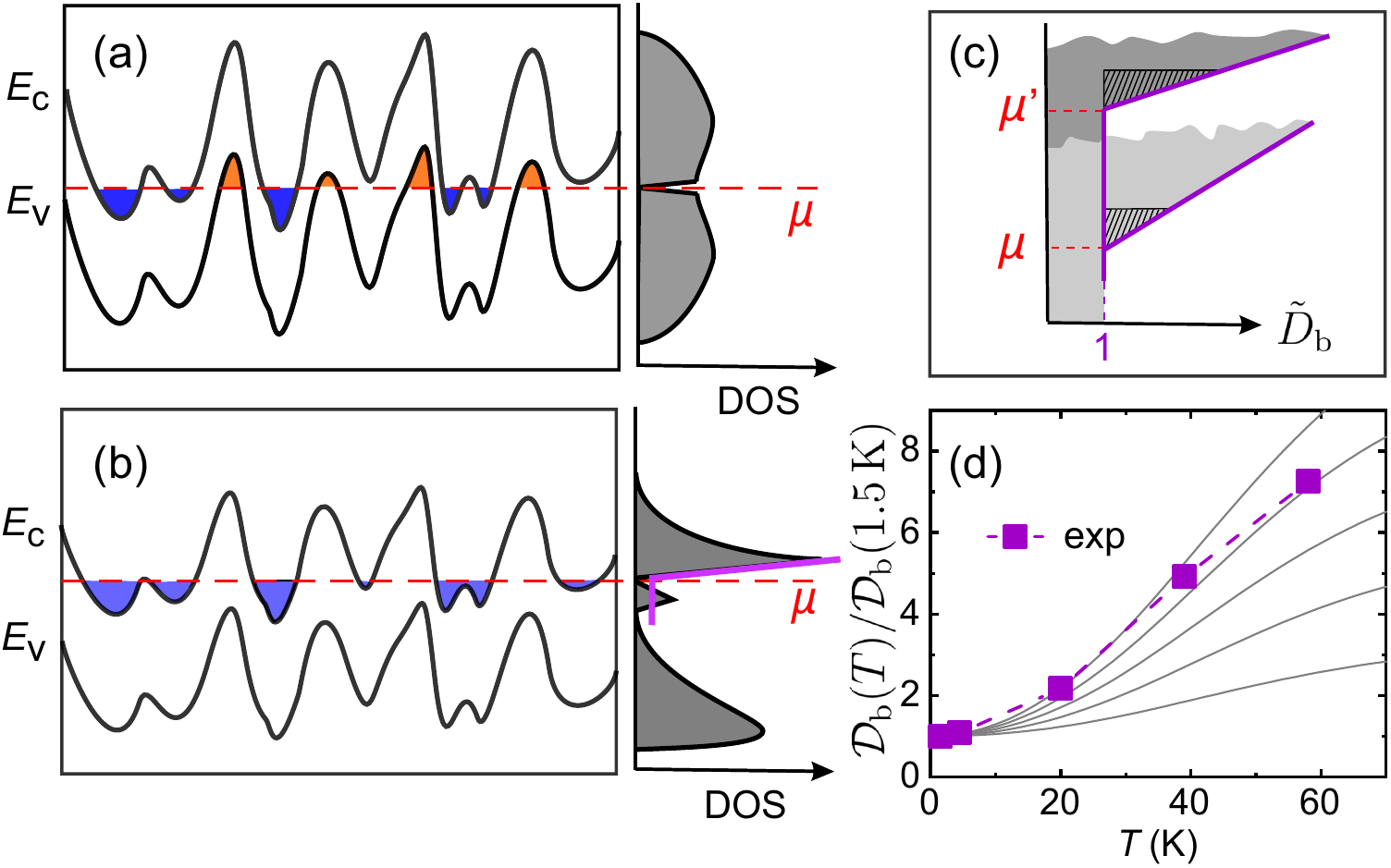}
	\caption{
		(a) Conduction ($E_\text{c}$) and valence band ($E_\text{v}$) profiles, fluctuating due to long range Coulomb interactions for perfect compensation. Electron- and hole puddles form. A sketch of the symmetric bulk DOS  including the Coulomb gap at $\mu$ is shown at right \cite{SkinnerPRL2012}. 
		(b) For non-perfect compensation, \eg \,for the donor concentration slightly higher than the one of acceptors, the fluctuation amplitude and the shape of the many particle DOS, sketched following Ref. \cite{SkinnerJETP2013} change considerably. We expect a similar situation for changing the gate voltage instead of compensation. 
		Note that the overall number of impurity states (shaded region at right) is fixed and independent of the shape. For modeling we use the purple shape of the DOS, ignoring the Coulomb gap at $\mu$.
		(c) DOS of our toy model mimicking the DOS of (b) (purple line). The hatched areas illustrate the origin of the TDOS temperature dependence. (d) Solid grey lines represent the calculated TDOS of our toy model with the parameter $a$ varying from 10 (bottom) to 50 (top) in steps of 10. Filled squares are experimental data.  All curves (data points) are normalized to their values at $T=1.5$\,K.
			} 
	\label{fig_puddles}
\end{figure} 
which in particular can massively reshape itself when $\mu$ is varied \cite{SkinnerJETP2013},
see Fig.~\ref{fig_puddles}(a)-(b).
The reshaping is a complex many-body problem and can be very slow \cite{amir2011,ovadyahu2013}.
The situation is further complicated by the presence of the Dirac surface states encasing the bulk.
Since precise timescales for BiSbTeSe$_2$ are not known, and a comprehensive theory covering all
aspects of our 3D TI scenario is not available, we can only argue along phenomenological lines. 
First, the timescale of about 1 minute needed to produce each data point after changing $V_{\text{tg}}$ 
is assumed sufficiently long for the reshaping to take place, at least partially.
Second, we look for a bare-bone DOS toy model meeting three fundamental constraints:
(i) the resulting TDOS is everywhere constant but strongly $T$-dependent;
(ii) the overall number of charges lost to the bulk when scanning $V_\text{tg}$ from one Landau gap
to the other increases by a factor of roughly 8 in the interval $T=1.5$\,K to $T=58$\,K;
(iii) its shape is qualitatively compatible with established theoretical results \cite{SkinnerJETP2013}.
Consider therefore the TDOS 
\begin{equation}
\label{bulk_TDOS}
\Tdos_{\rm b}(\mu) = \int \partial_\mu \left[D_{\rm b}(E,\mu) f(E-\mu,T)\right] {\rm d}E, 
\end{equation}
the $\mu$-derivative acting on both the Fermi function $f$ and $D_{\rm b}$.  The background DOS is given by a toy model:
\footnote{We do not consider an explicit $T$-dependence of the DOS as too little precise knowledge
is available for a meaningful guess.  Note also that we ignore the presence of a Coulomb gap \cite{EfrosShklovskii1984, grunewald1982, CoulombGapTheory, GeCoulombGapPRL2001, vaknin2002, bardalen2012, meroz2014}.
This is because the gap is a function of $E-\mu$, not of $\mu$ alone, 
and thus its contribution to $\partial n/\partial\mu$ can be neglected, at least within our phenomenological
approach (see the Supporting Information).}
\begin{equation}
\label{bulk_DOS}
\tilde{D}(E,\mu)\equiv \frac{D_{\rm b}(E,\mu)}{D_0} =
\left\{
\begin{array}{ll}
1 & {\rm if}\quad E_m^* < E \leq \mu,
\\
\left[1+a\mu(E-\mu)\right] & {\rm if}\quad \mu < E < E_M^*. 
\end{array}
\right.
\end{equation}
Here $D_0$ is the TDOS value measured at $T=0$, $a$ is a parameter, while $E_m^*, E_M^*$ are cut-off energies
such that $|E_m^*-\mu_1|, |E_M^*-\mu|\gg k_{\rm B}T_{\rm max}$, with $\mu_1$ the position of the lower Landau gap
and $T_{\rm max}=58$\,K (see corresponding sketch in Fig. S11 in the Supporting Information).  Beyond such cut-offs the form of $D_{\rm b}(E,\mu)$ is irrelevant for computing the corresponding TDOS. 
The dimensionless $\tilde{D}_{\rm b}(E,\mu)$ is sketched in Fig.~\ref{fig_puddles}(c) for two different values of $\mu$.
Notice that such a DOS is the result of a reorganization of impurity states, not of the appearance of additional states. That is, it is constrained by the condition that the overall number of donors and acceptors states within the gap is fixed. Thus, its profile above $\mu$ sharpens because impurity states from higher energies migrate closer to the chemical potential as the latter increases, while it stays flat below $\mu$ as it represents only the average value of the DOS in the region $E_m^* < E \leq \mu$ (see sketch in Fig.~\ref{fig_puddles}(c), and Fig. S11 in the Supporting Information). In other words, consider two values of the chemical potential $\mu'>\mu$, both within the energy region defined by the position of the lower ($\mu_1$) and upper Landau gap ($\mu_2$): The DOS reorganization is such that more higher energy states are available within $k_{\rm B}T$ of $\mu'$ than within $k_{\rm B}T$ of $\mu$, see hatched areas in Fig.~\ref{fig_puddles}(c), and this difference is responsible for a strongly $T$-dependent TDOS (see the Supporting Information). Thus, the overall amount of charges lost to the bulk between $\mu=\mu_2$ and $\mu=\mu_1$ is 
\be
\delta n_\text{b} = \int D_{\rm b}(E,\mu_2) f(E-\mu_2,T) {\rm d}E - \int D_{\rm b}(E,\mu_1) f(E-\mu_1,T) {\rm d}E.
\ee
As discussed above, this quantity \textit{cannot} be written as $\int\,D^{\rm rigid}_{\rm b}(E)
[f(E-\mu_2)-f(E-\mu_1)]{\rm d}E$, in terms of a rigid single-particle DOS $D^{\rm rigid}_{\rm b}(E)$.
In Fig.~\ref{fig_puddles}(d) we compare the measured TDOS normalized to its $T=1.5$\,K value, $\Tdos_{\rm b}(T)/\Tdos_{\rm b}(1.5\,{\rm K})$, and the one computed from Eqs.~\eqref{bulk_TDOS} and \eqref{bulk_DOS}.
Qualitatively, the similarity is evident.  We emphasize however that our toy model can only be taken as an empirical guide to the data.

\textit{Conclusions} -- By probing the capacitance of a BiSbTeSe$_2$ capacitor structure we are able to extract the electronic DOS as a function of the gate voltage (chemical potential $\mu$). Our experimental data, together with the calculations, show that the filling (via field-effect) of Dirac surface states and conventional bulk states is closely intertwined. While we observe the Landau quantization of the Dirac surface states in the quantum capacitance signal, the position of the Landau gaps and their increased separation on the gate voltage scale with increasing temperature can only be understood by considering bulk states. These experiments provide a so far unknown method to investigate the many-particle DOS of the bulk of a highly compensated TI. We find the corresponding bulk TDOS to be constant as a function of $\mu$, but strongly temperature dependent. This result is incompatible with a single-particle picture, and is evidence of the many-body character of the bulk phase. Indeed, the in-depth analysis of quantum capacitance data
suggests that the background density of states reorganizes whenever the gate voltage (chemical potential) is changed, on the time scale of a minute.  This is compatible with the slow (glassy) dynamics of a disordered and strongly interacting phase \cite{ovadyahu2013,meroz2014}, as expected in the bulk of a compensated TI \cite{SkinnerJETP2013,AndoPRB2016_OpticalConduc.Puddles,KnispelPRB2017}.
To the best of our knowledge, the dynamics of such a surface-bulk two-phase hybrid system is largely uncharted ground at the moment.  However a proper understanding of it is crucial for potential device concepts based on the properties of topological surface states.

\begin{acknowledgements}

We thank F. Evers for inspiring discussions, and G. Vignale for a careful reading of the manuscript. The work at Regensburg was funded by the Deutsche Forschungsgemeinschaft (DFG, German Research Foundation) - Project-ID 314695032 - CRC 1277 (Subprojects A07, A08). This project has received further funding from the European Research Council (ERC) under the European Union’s Horizon 2020 research and innovation programme (grant agreement No 787515, ProMotion), as well as the Alexander von Humboldt Foundation. The work at Cologne was funded by the Deutsche Forschungsgemeinschaft (DFG, German Research Foundation) - Project number 277146847 - CRC 1238 (Subproject A04).

\end{acknowledgements}



%

\cleardoublepage


\def\be{\begin{equation}}
	\def\ee{\end{equation}}
\def\ber{\begin{eqnarray}}
	\def\eer{\end{eqnarray}}
\def\nn{\nonumber}


	
\begin{center}
	\textbf{\Large{Supplementary information:
			2D-Dirac surface states and bulk gap probed via quantum capacitance in a 3D topological insulator}}
\end{center}

\begin{center}
	Jimin Wang,$^1$ Cosimo Gorini,$^2$ Klaus Richter,$^2$ Zhiwei Wang,$^{3,4}$ Yoichi Ando,$^3$ and Dieter Weiss$^1$\\
	\it{\small{$^1$Institute of Experimental and Applied Physics, University of Regensburg, 93040 Regensburg, Germany}}\\
	\it{\small{$^2$Institute of Theoretical Physics, University of Regensburg, 93040 Regensburg, Germany}}\\
	\it{\small{$^3$Physics Institute II, University of Cologne, Z\"ulpicher Str. 77, 50937 K\"oln, Germany}}\\
	\it{\small{$^4$Key Laboratory of Advanced Optoelectronic Quantum Architecture and Measurement, Ministry of Education, School of Physics, Beijing Institute of Technology, Beijing, 100081, China}}
\end{center}

\setcounter{equation}{6}
\setcounter{figure}{0}

\maketitle

\section{Details of sample fabrication, transport and capacitance measurements}

\label{C_measurements}
Pristine, slightly $p$-doped BiSbTeSe$_2$ crystals grown by the modified Bridgman method \cite{Ren2011s} were used. Angle-resolved photoemission spectroscopy has shown that Dirac point and $\mu$ of this material lie in the bulk gap \cite{AndoNC2012_ARPES_BSTSs}. BiSbTeSe$_2$ flakes were exfoliated onto highly $p$-doped Si chips (used as backgate) coated by 285 nm SiO$_2$. The flakes were processed into quasi-Hallbars with Ti/Au (10/100 nm) Ohmic contacts. A larger h-BN flake, transferred on top of BiSbTeSe$_2$ serves as gate dielectric. Finally, Ti/Au (10/100 nm) was used as a top gate contact. 
Transport measurements using low ac excitation currents (10 nA at 13 Hz) were carried out above 1.5 K and in a cryostat (Oxford instrument) with out of plane magnetic fields $B$ up to 14 T. 

The capacitance is measured in a straightforward and simple manner using an Andeen Hagerling 2700A capacitance bridge. As we want to measure the capacitance between the gate and the TI layer (Fig.~S1a), the gate is connected to the low contact (L) of the bridge and the TI layer is connected to the high contact (H). We were working in the bridge mode where for every data point (capacitance and loss) the bridge is balanced against the device under test (Fig.~S1b). The bridge is operating in the frequency range between 50 Hz and 20 kHz; we are using the lowest frequency in order to avoid resistive effects.\cite{KozlovPRL2016_QCs} )

\makeatletter 
\renewcommand{\thefigure}{S\@arabic\c@figure}
\begin{figure}
	\includegraphics[width=.5\linewidth]{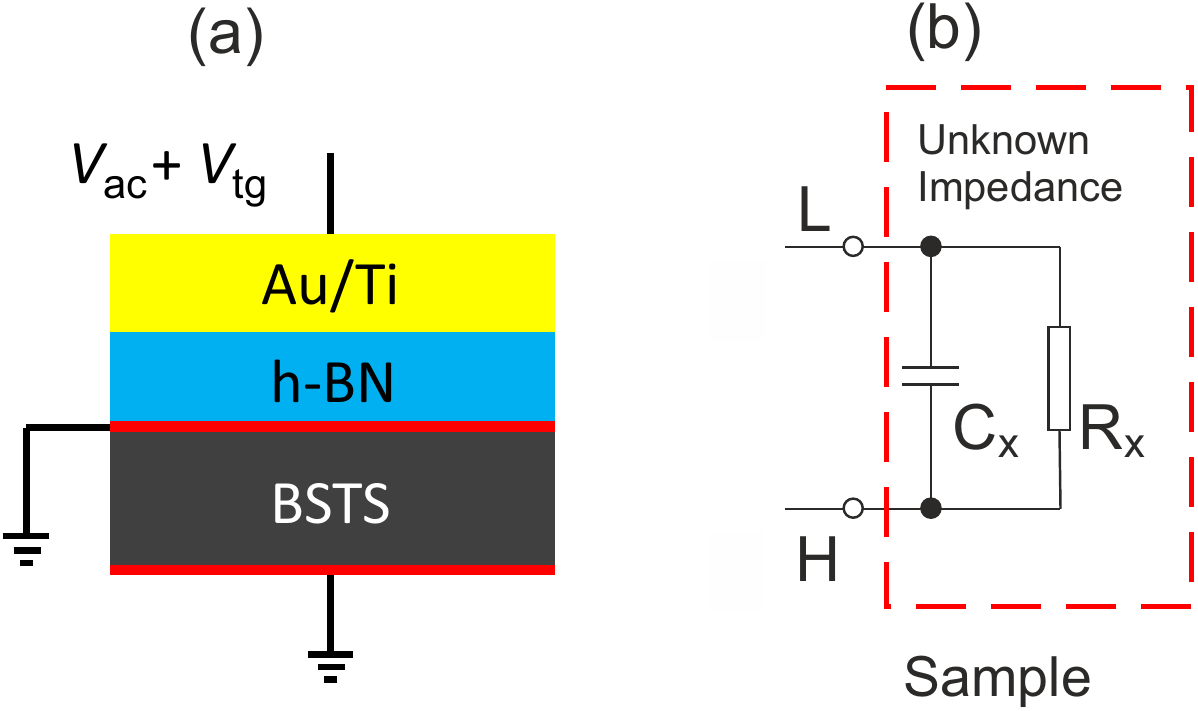}
	\caption{Capacitance measurement setup. (a) Measurement configuration, a small AC modulation voltage is superimposed on $V_\text{tg}$ to measure capacitance and loss. (b) Measurement principle.  } 
	\label{Figure_Cmeasurements}
\end{figure}
\makeatother

\section{Resistivity-temperature relation}
\label{RT}

Fig.~S2 shows the temperature dependence of the device discussed in the main part of the text. With decreasing temperature, the resistivity of the device increases down to $\sim65$ K, characteristic for semiconductor-like, bulk dominated conduction, and decreases below $\sim65$ K, typical for metallic-like, surface dominated conduction. 

\makeatletter 
\renewcommand{\thefigure}{S\@arabic\c@figure}
\begin{figure}[hbt!]
	\includegraphics[width=0.9\linewidth]{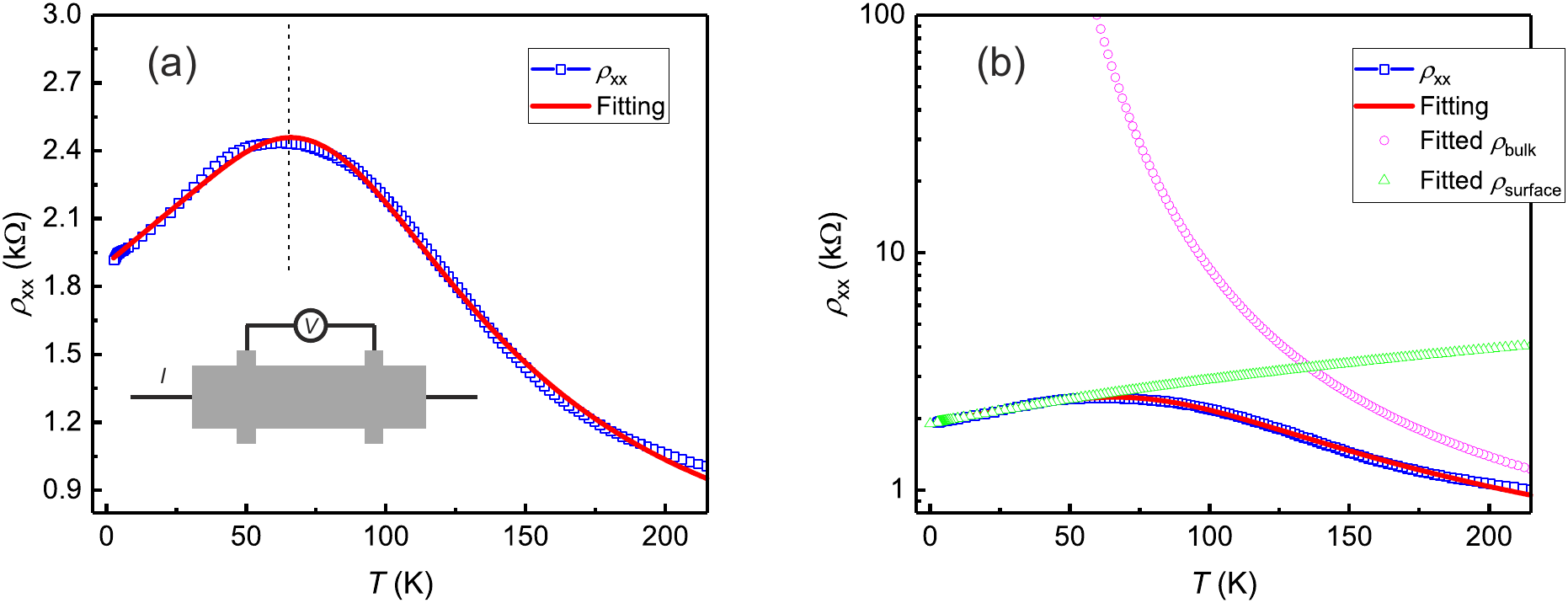}
	\caption{	$R$-$T$ relation of the device. (a) The experimental data (blue squares) and corresponding fit (red line). The dashed line approximately separates the bulk and surface conduction dominated regions. (b) The same graph as in (a) in log scale and together with the fitted (see text) bulk (purple) and surface (green) contributions. The inset in (a) shows the measurement configuration.} 
	\label{Figure_RT_curve}
\end{figure}
\makeatother

Useful information, such as carrier density $n$ and activation energy $E_\text{a}$ can be extracted from the graph. Following a recent paper \cite{Cai2018s}, the surface and the bulk of the topological insulator contribute to the conduction independently, equivalent to resistors connected in parallel. The resistivity of each surface ($\rho_\text{s}$) can be modeled as metallic like, $\rho_\text{s}=\rho_\text{s,0}+A\cdot T$, where $\rho_\text{s,0}$ and $A$ are constants. The bulk resistivity $\rho_\text{b}$ is thermally activated for temperatures beyond the variable range hopping regime: $\rho_\text{b}=\rho_\text{b,0}\cdot {\rm exp}(E_\text{a}/k_\text{B} T)$, where $\rho_\text{b,0}$ is a constant, $E_\text{a}$ is half the energy gap $\Delta$, and $k_\text{B}$ is the Boltzmann constant. Thus the total resistivity is $\rho = (2\cdot \rho_\text{s}^{-1} + \rho_\text{b}^{-1})^{-1}$. Here we assume, for simplicity, that the two surfaces are equivalent, having the same carrier density $n_\text{s}$ and mobility $\mu_\text{s}
$. The corresponding fit is shown in Fig.~S2a, where we obtained remarkable good agreement. In addition, the 
fitting also yields $\rho_\text{s}=3801+20.5\cdot T$, and $\rho_\text{b}=234\cdot {\rm exp}(31\ \text{meV}/k_\text{B} T)$ and thus the effective energy gap $\Delta^{'}=62\,\text{meV}$. Compared to the nominal energy gap $\Delta \approx 300\,\text{meV}$\cite{AndoNC2012_ARPES_BSTSs}, the effective gap is 5-times smaller. This is consistent with previous reports\cite{SkinnerPRL2012s} and due to thermal excitation of electrons from the Fermi-level to the percolation energy which is much closer to $\mu$ than the band edge.

At sufficiently low temperature ($e.g.$, $1.5$ K), the conduction is almost entirely dominated by Dirac surface states, also verified in Ref. \cite{XuNatPhys2014s}. However, at higher temperatures, bulk carriers are activated and contribute to transport.  At $T$ = 58\,K, using the results from the fit above, and by assuming a typical, temperature independent value of $\mu_\text{b}=200\,$cm$^2$/Vs, we obtain $n_\text{b}=2.7\cdot 10^{11} $cm$^{-2}$. The same way, the carrier density at zero gate voltage on each surface is $n_{\rm s}=6.3\cdot 10^{11}$\,cm$^{-2}$, using $\mu_{\rm b}=2000\,$cm$^2$/Vs. Thus the delocalized bulk carrier density at high $T$ is comparable to that in the surface states.

\section{Determining $\mathbf{\Delta V_{tg}}$}
\label{Delta_V}

Here we show how $\Delta V_\text{tg}$ in Fig.\,2 and Fig.\,3 is determined. As an example we use the experimental data taken at 20 K and 14 T, shown in Fig.~S3. To enhance the visibility of the minima and to suppress effects of the background we calculate the  second derivative of the trace. This results in two pronounced peaks at the position of the minima. The distance between the 2 peaks gives $\Delta V_\text{tg}$. This procedure is particularly useful at elevated temperatures where it is more difficult to determine the minimum position accurately from the original capacitance trace.

\makeatletter 
\renewcommand{\thefigure}{S\@arabic\c@figure}
\begin{figure}
	\includegraphics[width=.7\linewidth]{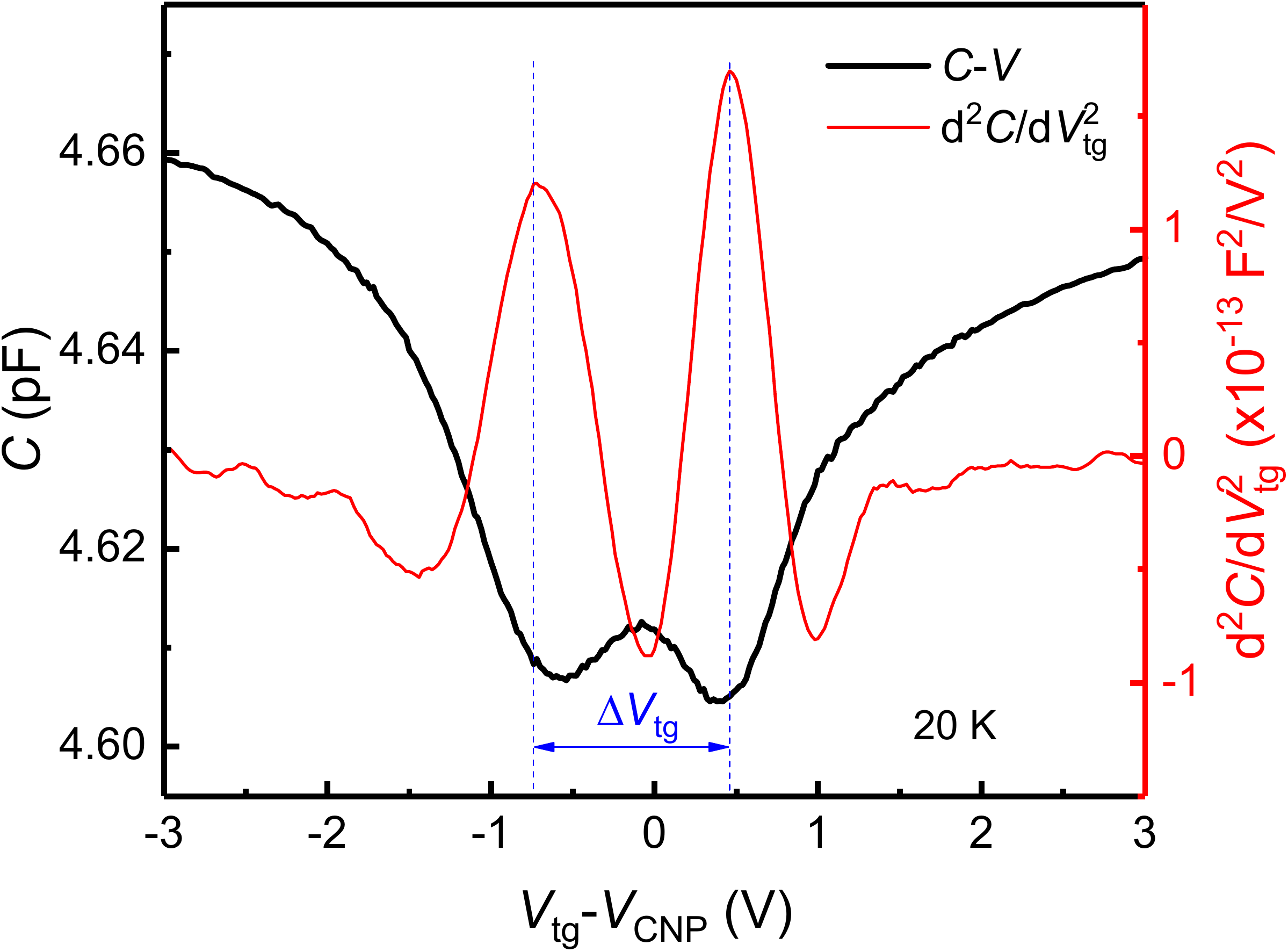}
	\caption{\textit{C-V} trace at 20 K and 14 T, and the corresponding second derivative $d^2C/dV_{\rm tg}^2$, used to determine $\Delta V_\text{tg}$.}
	\label{Figure_Determine_Delta_V}
\end{figure}
\makeatother

\section{Failure of fitting \textit{C-V} trace at 14 T without background DOS}

\label{bad fitting witout Db}
In Fig.~S4 we compare the capacitance measured at $B=14$\,T and $T =1.5$\,K to fits with and without using a constant background density. The fits without background density fail to describe the data, especially the voltage difference between the capacitance minima which correspond to the Landau gaps. The distance between calculated, adjacent Landau gaps is always smaller (marked by $\Delta V_\text{tg}$ in Fig.~S4) than observed in experiment if $D_\text{b}$ is set to 0, no matter how the other parameters are chosen. This discrepancy becomes much larger at higher temperatures.

\makeatletter 
\renewcommand{\thefigure}{S\@arabic\c@figure}
\begin{figure}
	\includegraphics[width=0.8\linewidth]{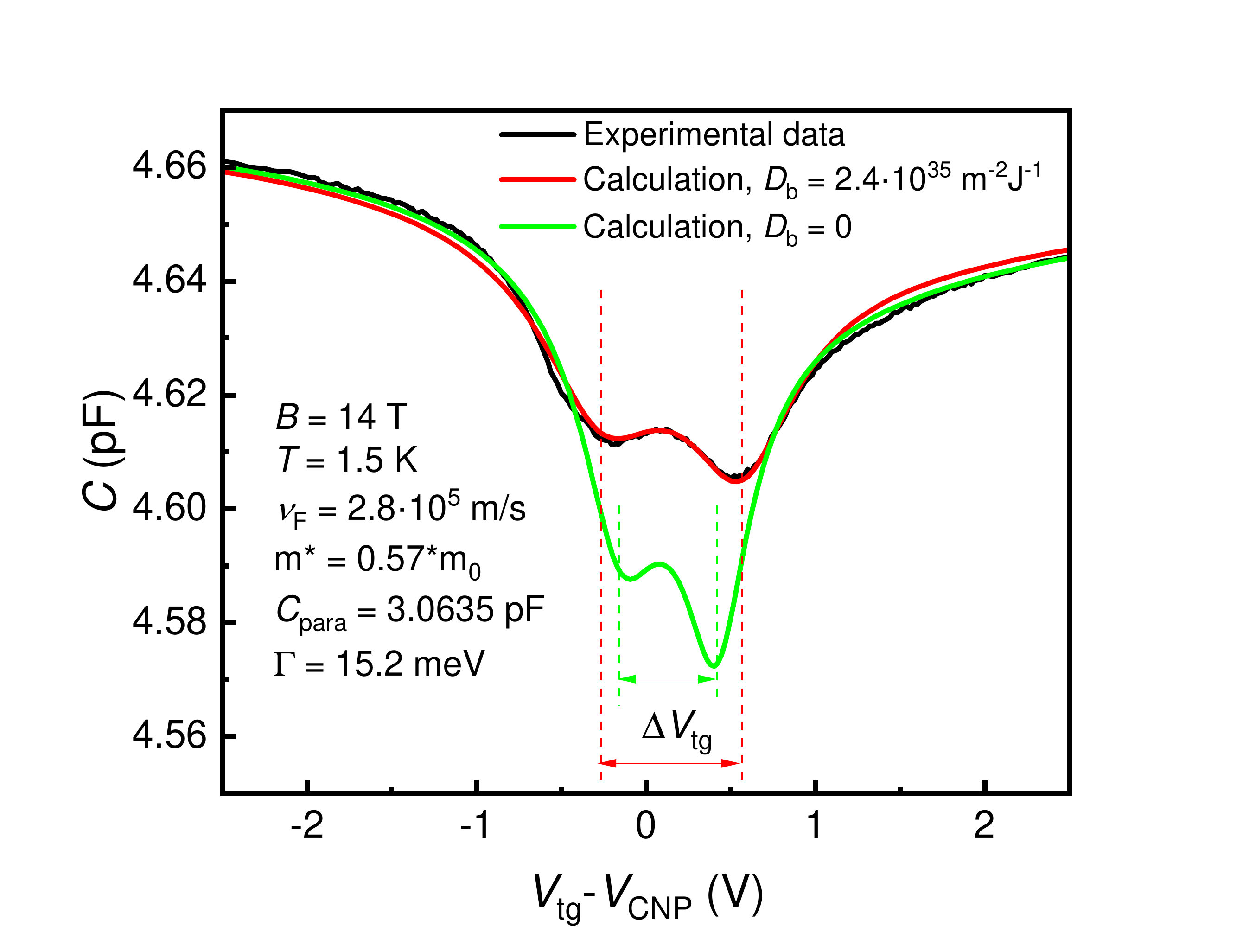}
	\caption{Comparison of fits to the experimental data at 1.5 K and 14 T, with and without finite $D_\text{b} = 2.4\cdot10^{35}\ \text {m}^{-2}\text {J}^{-1}$, while the other paramters are kept the same: Fermi velocity $v_{\rm F}=2.8\cdot10^{5}\ \text{m/s}$, effective mass $m^{*}= 0.57 m_0$, LL broadening (Gaussian) $\Gamma = 15.2\ \text{meV}$ and parasitic capacitance $C_\text{para}= 3.0635\ \text{pF}$.} 
	\label{Figure_badfitting}
\end{figure}
\makeatother

\section{Transport data and estimating the filling rate from Hall measurements}

\label{Transport results}
Fig.~\ref{Figure_TransportDetails}a and Fig.~\ref{Figure_TransportDetails}b show $\rho_{\rm xx}$ and $\rho_{\rm xy}$ vs $V_\text{tg}$ at $V_\text{bg}$ = 0 V and different $B$. Note, that in transport top and bottom surfaces contribute to the signal while in capacitance experiments we probe preferentially the top surface. Therefore, capacitance and transport data cannot directly be compared. The data shown here are consistent with Fig. 1c in the main text.  

The filling rate $\text dn/\text dV_\text{tg}$ extracted from the position of the Landau gaps on the gate voltage scale in the main text (Fig.\,2(a) and Fig.\,3(a)) is smaller than $C_0/(Ae)$, expected from the insulator capacitance $C_0$, $A$ is the area of the capacitor. Fig.~\ref{Figure_TransportDetails}c shows that the reduced filling rate is also measured by the classical Hall slope.

The Hall data were taken at 1.5 K by sweeping the top gate voltage ($V_\text{tg}$) at fixed $B$, as well as by sweeping $B$ at fixed $V_\text{tg}$, at $V_\text{bg}$ = 0 V. The device shows at such low temperature surface dominated conduction, which takes place in top and bottom surface.  From Fig.\,1(c) in the main text, we see that at zero gate voltage, both top and bottom surfaces are slightly \textit{p}-doped. By grounding the back gate and sweeping $V_\text{tg}$ from 5 to -5 V, one changes the carrier type in the top surface. For  $V_\text{tg} < 0$ in Fig.~\ref{Figure_TransportDetails}c, \textit{p}-type conduction prevails in top and bottom surfaces. The total carrier density and the total filling rate in this regime can be simply obtained from the one carrier Drude model using the linear Hall slope. The filling rate, \textit{i.e.}, the change of total density with gate voltage, is given by the slope of the red line in Fig.~\ref{Figure_TransportDetails}c.  The corresponding filling rate is $3.2\cdot10^{11}\ \text{cm}^{-2}\text{V}^{-1}$.  This is consistent with the filling rate displayed by the QHE data in the main text ($4\cdot10^{11}\text{cm}^{-2}\text{V}^{-1}$, using $\Delta V_{\rm tg} =0.84$\,V from Fig.\,3(c)). Similar conclusions can be found in \cite{XuNatPhys2014s}.

\makeatletter 
\renewcommand{\thefigure}{S\@arabic\c@figure}
\begin{figure}
	\includegraphics[width=1\linewidth]{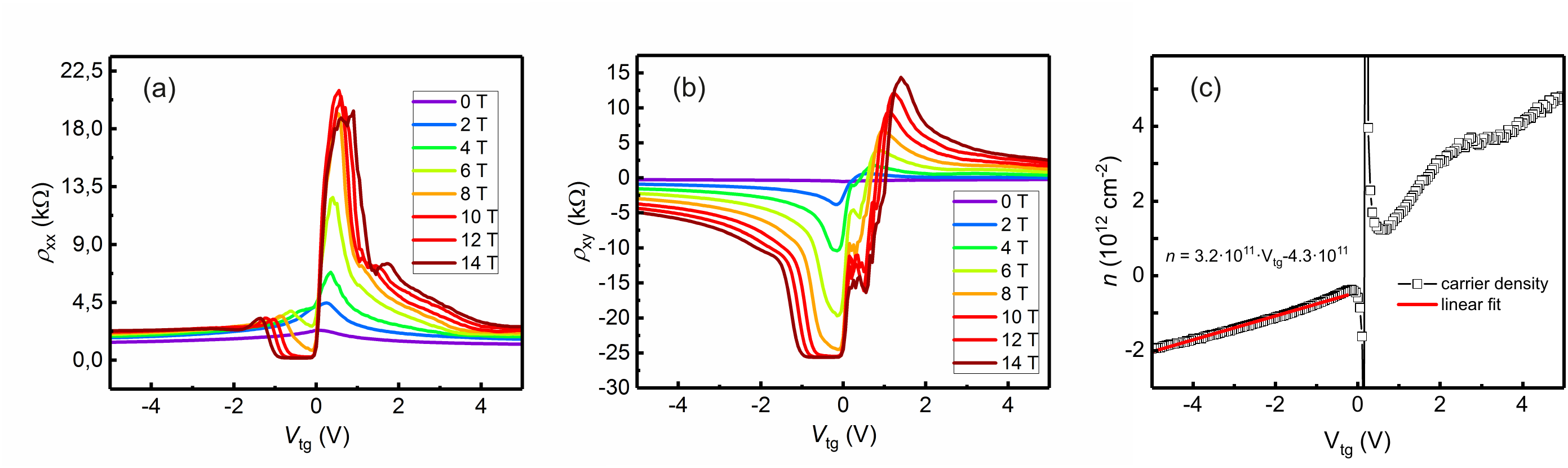}
	\caption{Magnetic transport results at $V_\text{bg}$ = 0 V. (a) $\rho_{\rm xx} (V_\text{tg})$ traces for different $B$. (b)$\rho_{\rm xy} (V_\text{tg})$ traces for different $B$. (c)Estimate of the total carrier density \textit{p} using the Hall slope giving $p=(e\frac{\text{d}\rho_\text{xy}}{\text dB})^{-1}$ for low magnetic field data. Here the electron density is defined positive, the hole density negative. The extraction of filling rate is reliable for the hole side, where both top and bottom surfaces have the same carrier type. It is not reliable on the electron side, where the two surfaces have different carrier types. } 
	\label{Figure_TransportDetails}
\end{figure}
\makeatother
\section{Comparing fitting of \textit{C-V} at $B = 0$ T and $T = 1.5$ K with and without background DOS}

\label{fitting at B0T,T1.5K with and without Db}
The $B = 0$ capacitance trace in Fig.\,1(d) of the main text we fitted without using a background DOS $D_\text{b}$. While the $B = 0$ data can be well fitted using $D_\text{b} = 0$, the capacitance data in quantizing magnetic field cannot. In Fig.~\ref{Figure_B0T1.5D_b} we show that the data of Fig.\,1(d) can be likewise fitted with a finite DOS $D_\text{b}$. As a result, the energy broadening becomes significantly smaller, but the important parameters like Fermi velocity $v_{\rm F}$ and effective mass $m^*$ stay within 20\% the same.

\makeatletter 
\renewcommand{\thefigure}{S\@arabic\c@figure}
\begin{figure}
	\includegraphics[width=.9\linewidth]{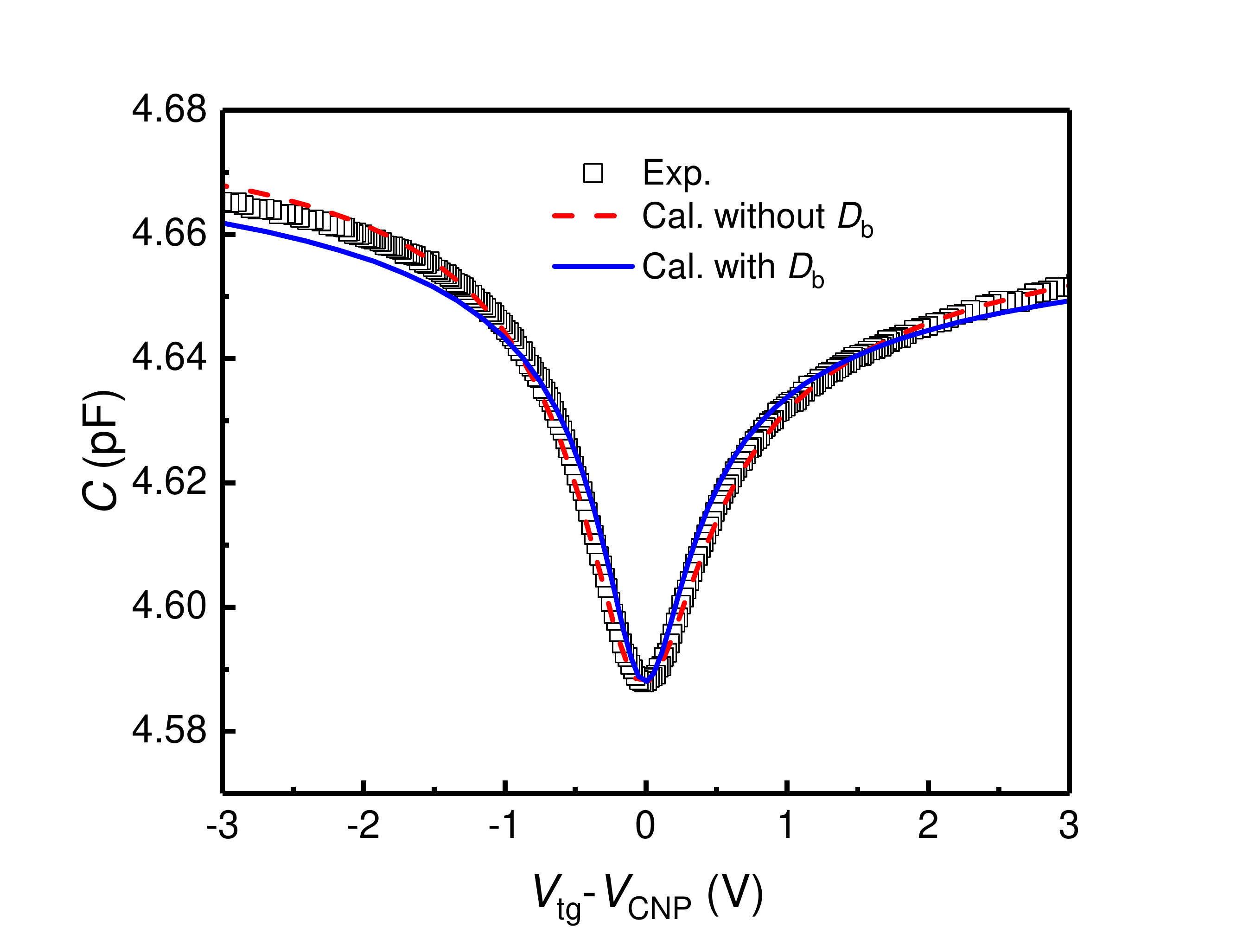}
	\caption{Fitting the data of Fig.\,1(d)  with and without $D_\text{b}$ at $B$ = 0\,T and $T$ = 1.5\,K. The red dashed line is the same as the one shown in Fig.\,1(d), with fitting parameters  $v_{\rm F} =3.2\cdot 10^{5}$\,m/s, $m^*=0.47 m_0$, $\sigma = 29.4$ meV, parasitic capacitance 3.073 pF, and $D_{\rm b} = 0$. The blue trace, in contrast, was obtained using the same $D_{\rm b} = 2.4 \cdot 10^{35}\, \text{m}^{-2}\text{J}^{-1}$ extracted at the same $T$ from the 14\,T data. The other fit parameters are then  $v_{\rm F} =2.8\cdot 10^{5}$\,m/s, $m^*=0.57 m_0$, $\sigma = 15.2$ meV, and parasitic capacitance 3.065\,pF. } 
	\label{Figure_B0T1.5D_b}
\end{figure}
\makeatother

\section{Extracting the quantum capacitance from $C(V)$ curves }
\label{Qc extraction}

Using Eq.(1) from the main text one can extract the quantum capacitance $C_{\rm q}=Ae^2D(\mu)$, which is given by
$C_{\rm q}= \frac{C(V)C_0}{C_0-C(V)}$. Here, $V$ stands for $V_{\rm tg}-V_{\rm CNP}$, used in the main text, and $C(V)=C_{\rm meas}-C_{\rm para}$, $A$ is the area of the capacitor. The extracted quantum capacitance depends strongly on the parasitic capacitance $C_{\rm para}$ which needs to be subtracted from the measured $C_{\rm meas}$ to get the real capacitance $C(V)$. $C_0$ is the geometric capacitance given by $C_{0}= \frac{\epsilon_{0}\epsilon A}{d}$, where $\epsilon$ and $d$ are dielectric constant and thickness of the gate insulator, respectively. Subtracting $C_{\rm para}$ and extracting $C_{\rm q}$ is thus prone to error. This is illustrated in Fig.~\ref{Figure_Compare}, where we show the extracted quantum capacitance per unit area ($C_{\rm q}/A$) using three slightly different (by 0.13\%) $C_{\rm para}$, both as function of gate voltage $V$ and chemical potential. 
\makeatletter 
\renewcommand{\thefigure}{S\@arabic\c@figure}
\begin{figure}
	\includegraphics[width=1\linewidth]{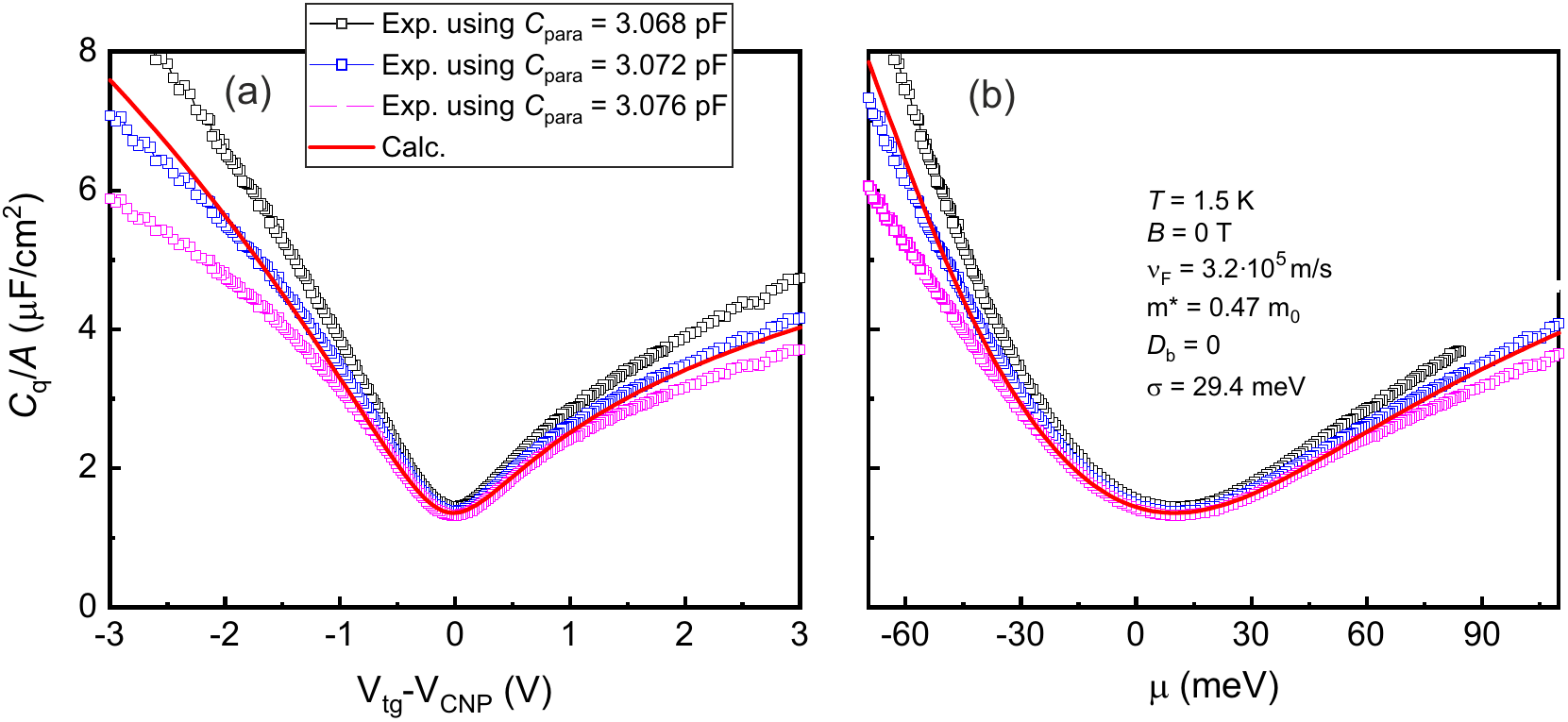}
	\caption{Quantum capacitance extracted for three values of the parasitic capacitance as a function of (a) gate voltage and (b) chemical potential. The red solid line is the model quantum capacitance used in Fig.\,1(d) to fit $C(V)$. It is consistent with the extracted $C_{\rm q}/A$, if $C_{\rm para}=3.072$\,pF is used.} 
	\label{Figure_Compare}
\end{figure}
\makeatother
Changing $C_{\rm para}$ a bit changes the subtracted $C_{\rm q}/A$ significantly. It is therefore much less critical to directly fit the capacitance $C_{\rm meas}$, instead of extracting first the quantum capacitance $C_{\rm q}/A$ and then fit $C_{\rm meas}$. Due to the inaccurately known $C_{\rm para}$ this is - as pointed out above - prone to error and we cannot expect a reliable fit. In case we fit $C_{\rm meas}$ directly, as we have done, $C_{\rm para}$ is only a constant offset which shifts the capacitance value on the $y$-axis. After the capacitance has been fitted and $C_{\rm para}$ determined, the extracted quantum capacitance is equivalent to the one used to fit the capacitance (red solid curves in Fig.~\ref{Figure_Compare}).
\makeatletter 
\renewcommand{\thefigure}{S\@arabic\c@figure}
\begin{figure}
	\includegraphics[width=0.5\linewidth]{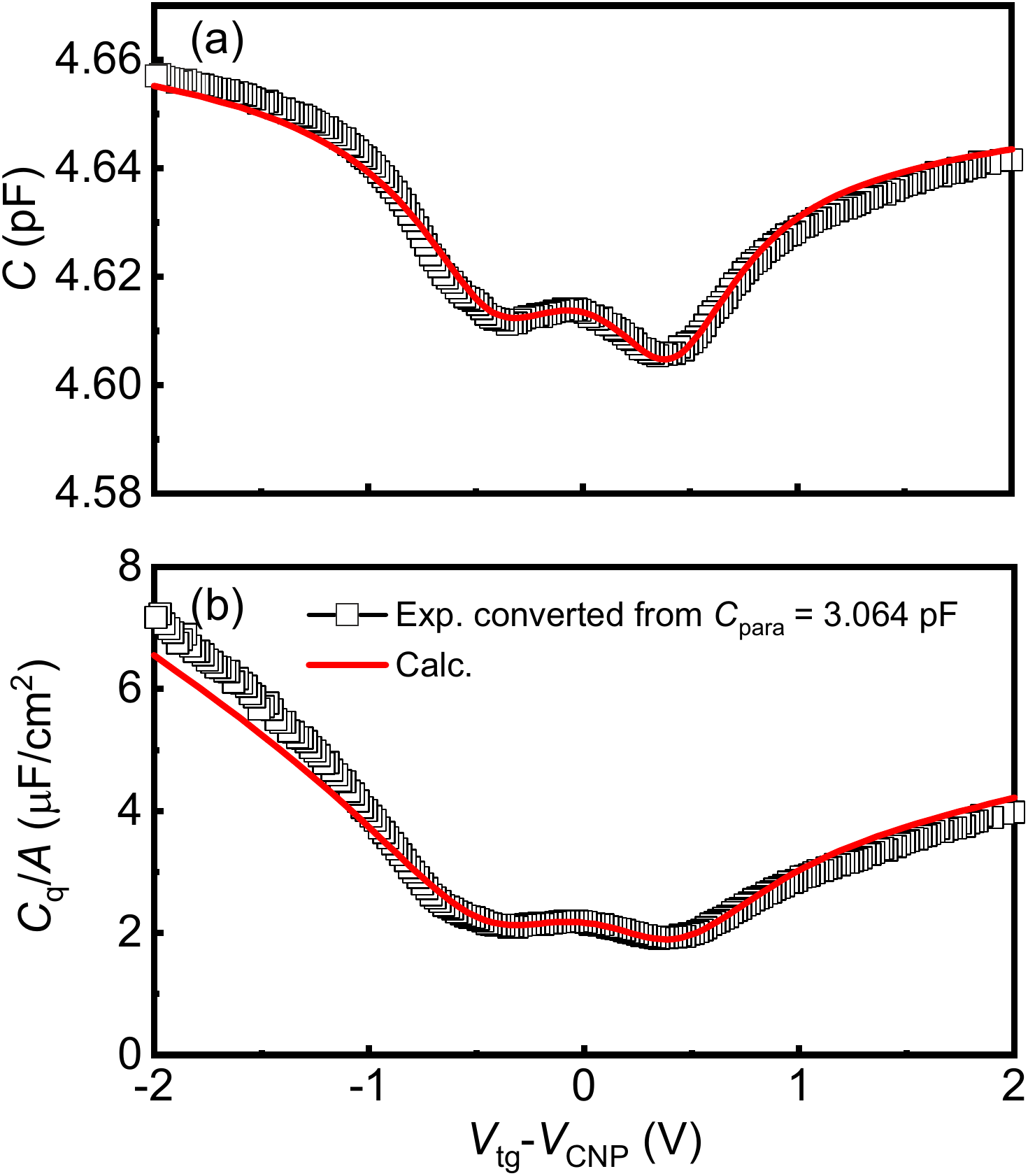}
	\caption{Fitting of data at $B = 14$\,T and $T = 1.5$\ K. (a) Measured total capacitance and corresponding fit. (b) Extracted quantum capacitance using a parasitic capacitance of 3.064\,pF. The red line is the quantum capacitance used to fit the capacitance in part (a). Parameters were $T = 1.5$\,K, $B = 14$\,T, $v_{\rm F}=2.8 \cdot 10^5$\,m/s, $m^*=0.57m_0$, $\Gamma = 15.2$\ meV, $D_{\rm b}=2.4\cdot 10^{35}$\,m$^{-2}$J$^{-1}$ and $C_{\rm para} = 3.064$\,pF. }
	\label{Figure_Compare2}
\end{figure}
\makeatother
The same holds for the high magnetic field data, shown in Fig.~\ref{Figure_Compare2}. We used the same procedure to extract the quantum capacitance from the 14\,T capacitance traces. In Fig.~\ref{Figure_Compare2} we compare capacitance and corresponding $C_{\rm q}/A$ on the same gate voltage scale. Both traces are well fitted by the same parameters. 

\section{Does the strong temperature dependence of the TDOS come from a single particle gap?}

Here we explore the possibility whether the temperature dependent TDOS might stem from a narrowed energy gap, which corresponds to the measured reduced activation energy of about 60 meV (see "1. Resistivity-temperature relation"). For testing purposes one can ad hoc construct a model DOS which mimics a reduced bandgap (here 60 meV) and adjust the value of the DOS such, that the \textit{average} TDOS between the two Landau gaps (-20 to 25 meV) corresponds, at high temperatures, to the one observed in experiment (our Fig. 3d). This average TDOS would cause the same shift of the capacitance minima on the gate voltage scale as observed in experiment. This is illustrated by Fig.~\ref{Figure_bandedge1}.

\makeatletter 
\renewcommand{\thefigure}{S\@arabic\c@figure}
\begin{figure}
	\includegraphics[width=1\linewidth]{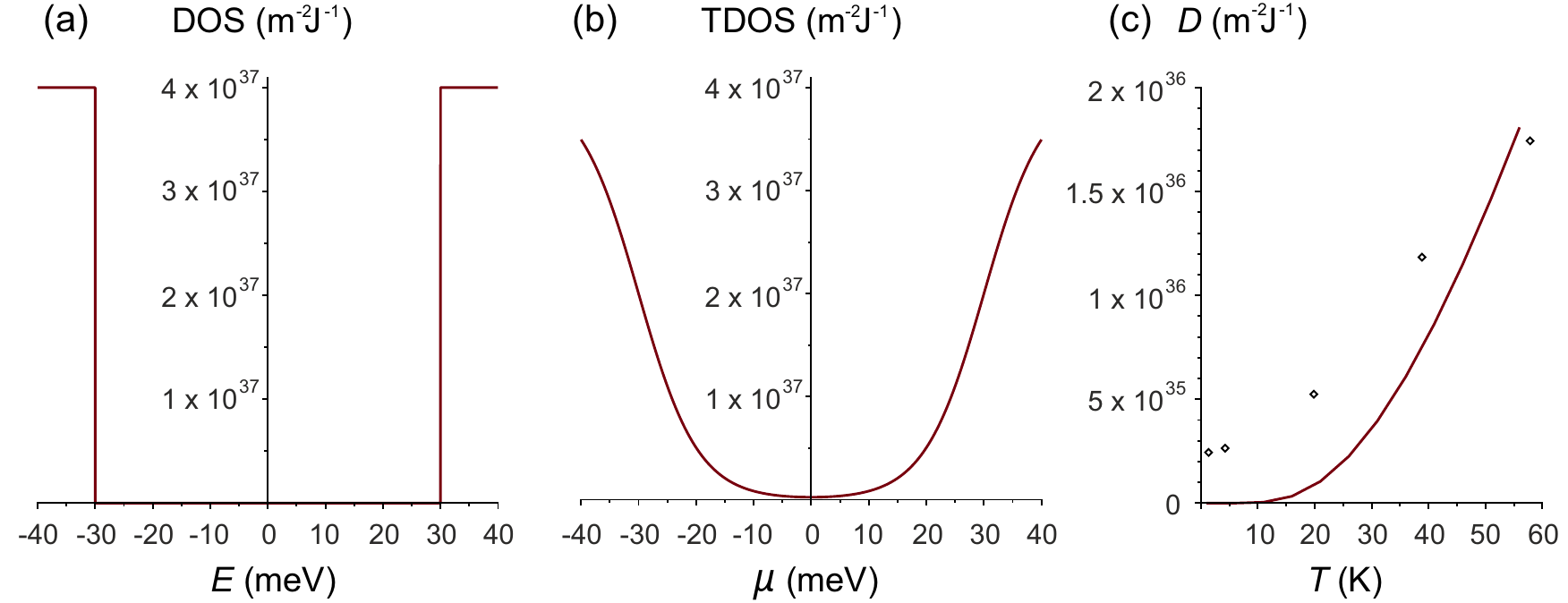}
	\caption{(a) Model DOS which mimics a reduced bandgap of 60 meV. (b) Corresponding TDOS calculated for T= 39 K. (c) Average TDOS as a function of T. The points are the extracted $D_{\rm b}$ from the main text (see Fig. 3d).}
	\label{Figure_bandedge1}
\end{figure}
\makeatother

Fig.~\ref{Figure_bandedge1} above shows (a) the corresponding DOS, (b) the TDOS and (c) the average TDOS, intended to mimic our background DOS. However, such a picture is not able to explain the finite DOS (TDOS) we find at the lowest temperatures ($T\textless\,20$ K). Even worse, the strong dependence of the TDOS on $\mu$ is inconsistent with experiment. This is shown in Fig.~\ref{Figure_bandedge2}, where we compare, as an example, the 39 K capacitance curve, calculated with a constant background DOS (as the one shown in Fig. 3b of our main text) with the capacitance we obtain from the 39 K TDOS of Fig.~\ref{Figure_bandedge2}b.

\makeatletter 
\renewcommand{\thefigure}{S\@arabic\c@figure}
\begin{figure}
	\includegraphics[width=.9\linewidth]{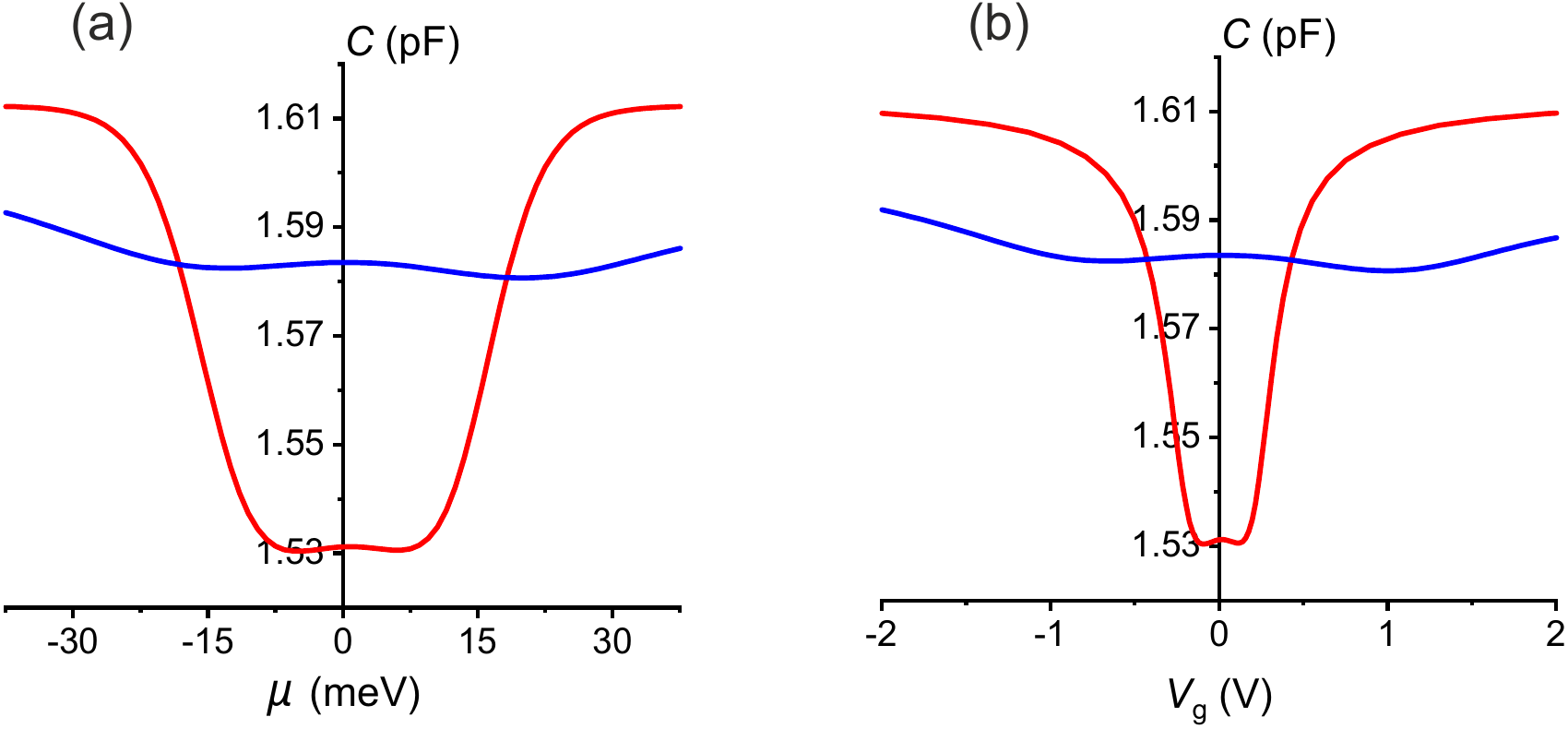}
	\caption{Capacitance as function of (a) energy and (b) voltage, calculated using a constant background $D_{\rm b}=11.8\cdot 10^{35}$\,m$^{-2}$J$^{-1}$ as used in the main text at 39 K (blue trace, see also Fig. 3d) compared to a capacitance trace where we assumed that the background comes from thermal smearing of the band edges. For the LL width we use $\Gamma$= 13 meV. Data are shown without parasitic capacitance which causes a constant capacitance offset.}
	\label{Figure_bandedge2}
\end{figure}
\makeatother

The strong dependence of the TDOS on $\mu$ completely dominates the capacitance signal so that the additional Landau structure is hardly visible. The shape of the trace is inconsistent with the experimental one. This picture does not change when we vary the parameters, increasing e.g. the “bandgap” and DOS step to keep the average TDOS constant. The density of states we observe in experiment is either nearly constant or has a weak dependence on $\mu$ and thus for the most part cannot come from states energetically far away from $\mu$.

\section{Details on the DOS toy model}
\label{toy_DOS}

We look for a minimal model for the effective DOS $D(E,\mu)$ qualitatively reproducing the TDOS data,
and compatible with known theory results coming from microscopic simulations \cite{SkinnerJETP2013s}.
To ease comparison with the latter we consider to be in the $n$-regime, but the construction would be analogous
in the $p$-regime.  Our model $D(E,\mu)$ is not $T$-dependent, as too little is known about it
with precision for a meaningful guess: On one hand, one expects from general arguments higher temperature to improve screening and thus reduce bulk potential fluctuations \cite{AndoPRB2016_OpticalConduc.Puddless}; On the other Dirac surface states provide some screening already at $T=0$ \cite{KnispelPRB2017s}.  Our toy model implicitly assumes that
$T$-induced screening is negligible compared to that from surface states and from charges added/removed by changing $\mu$.  

We start by splitting the TDOS so as to isolate its $T$-dependent part
\ber
\frac{\partial n}{\partial \mu} &=& \frac{\partial}{\partial \mu}\int\,{\rm d}E\, D(E,\mu)f(E-\mu,T)
\nn\\
&=&
\frac{\partial}{\partial \mu}\int\,{\rm d}E\, \underbrace{D(E,\mu)}_{D_\mu}
\left[ \underbrace{f(E-\mu,0)}_{f_0} + \underbrace{f(E-\mu,T)-f(E-\mu,0)}_{\delta f_\mu}\right]
\nn\\
&=&
\frac{\partial}{\partial \mu}\int\,{\rm d}E\, D_\mu f_0 + \frac{\partial}{\partial \mu}\int\,{\rm d}E\, D_\mu \delta f_\mu
\equiv
\frac{\partial n_0(\mu)}{\partial \mu} + \frac{\partial n_T(\mu,T)}{\partial \mu},
\eer
where $f$ is the Fermi function.  One has
\be
\delta f_\mu \approx \frac{\partial f}{\partial T} T = (E-\mu)\frac{\partial f}{\partial \mu}.
\ee
Whatever its form, the DOS has to respect the constraint
\be
\label{eq_normalization}
\int\,{\rm d}E\,D(E,\mu) = N = N_D + N_A\quad \forall\mu,
\ee
with $N_D, N_A$ respectively the donor and acceptor concentrations in the compensated TI.  However we are only concerned
with TDOS measurements in the small energy window $\Delta E\sim45$\,meV\, corresponding to the energy difference between the lowest two LL gaps, see Fig.~2 in the main text.  We thus need not worry about the DOS normalization condition, and assume the DOS to be of the following (dimensionless) form 
\be
\label{eq_toy_DOS}
\tilde{D}(E,\mu) = 
\left\{
\begin{array}{ll}
	1 & E_m^* < E \leq \mu, \\
	1 + a\mu(E-\mu) & \mu < E < E_M^*.
\end{array}
\right.
\ee
Here $a$ is a constant with dimensions ${(\rm eV)}^{-2}$, and $|E_m^*-\mu_1|,|E_M^*-\mu|\approx 20\,{\rm meV} \gg k_BT_{\rm max}$ ($T_{\rm max}=58$K) are energy cut-offs, with $\mu_1$ the position of the lower Landau gap(see sketch in Fig~\ref{fig:newmodel}). Physically, at $E > E_M^*$ the DOS shows features beyond the abrupt increase close to $\mu$
-- which is our only concern here.  Recall, however, that such an increase \textit{requires} the DOS to decrease
somewhere else, ensuring the overall normalization condition Eq.~\eqref{eq_normalization},
see Ref.~\cite{SkinnerJETP2013s}.  Similarly, our data shows that the (average of the) DOS below $\mu$ 
stays constant -- this means, \textit{inter alia}, that we cannot discern signatures of negative compressibility
\cite{eisenstein1992s,efros2008s} down to $T=1.5\,$K.  Its eventual increase at energies $E < E_m^*$ is irrelevant for computing $\partial n/\partial\mu$ away from such a region,
see Fig.~\ref{fig:newmodel} and corresponding caption.
The critical feature of Eq.~\eqref{eq_toy_DOS} is that its particle-hole asymmetry, determined by its increase for
$E>\mu$, depends directly on $\mu$.  This mimics the DOS reshaping, see Fig.~\ref{fig:newmodel}, 
and means that at a given $T$ more charges
can be loaded into the bulk at $\mu'$ than at a lower (electro-)chemical potential value $\mu<\mu'$, yielding $\partial n_T/\partial\mu\neq0$.
\makeatletter 
\renewcommand{\thefigure}{S\@arabic\c@figure}
\begin{figure}
	\includegraphics[width=.7\linewidth]{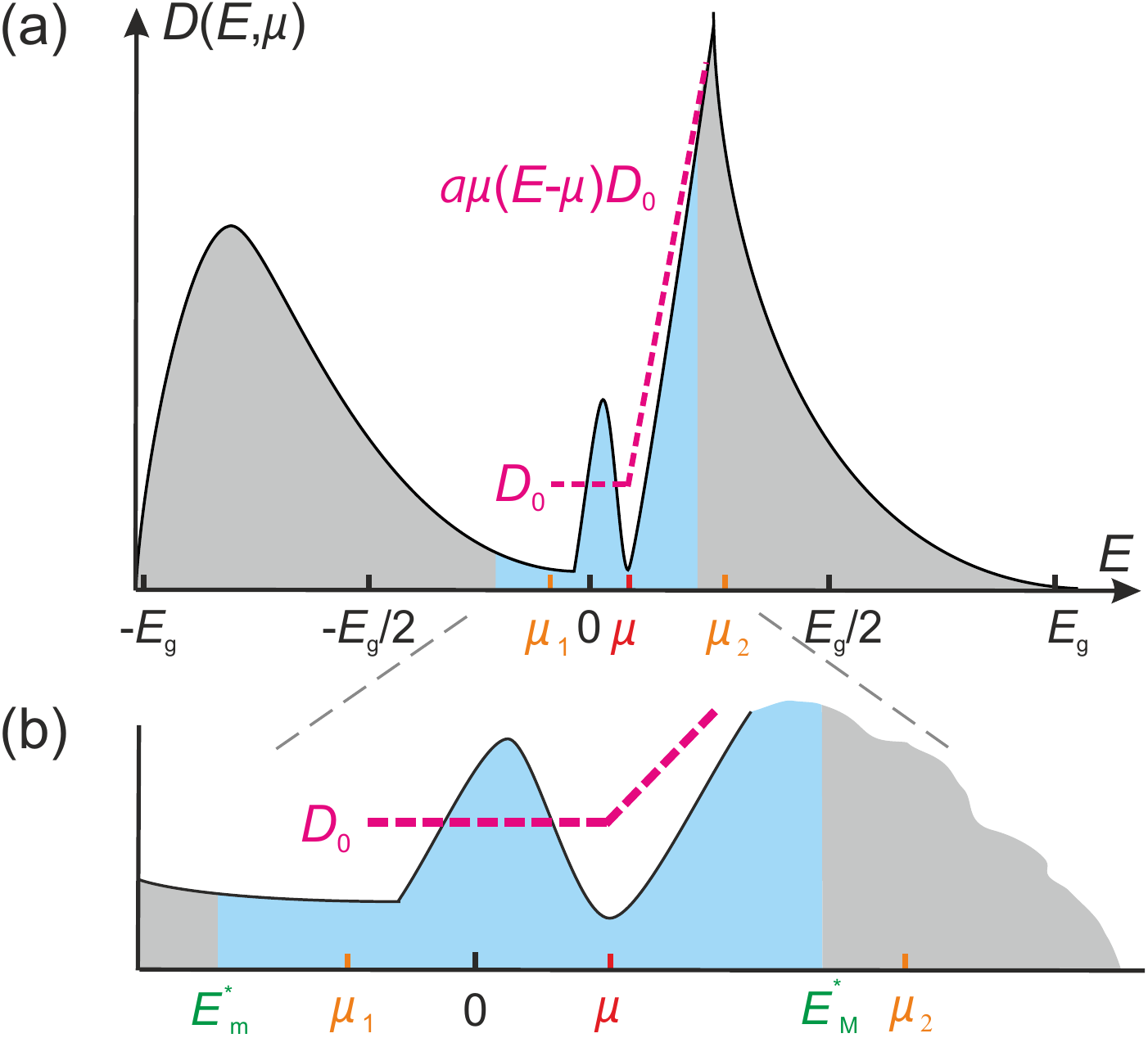}
	\caption{Sketch of the bulk DOS. The gap width is $E_g\approx 300$\,meV, while our conclusions are drawn from TDOS data in the comparably small energy window $\mu_2-\mu_1 \approx 45$\,meV. The region $E<E_m^*$ hosts a number $N_0$ of (almost exlusively) filled acceptor states, ergo a corresponding charge $n_0$. This charge is unknown, but also irrelevant for determining the TDOS $\partial n/\partial\mu$ in the region $E_m^* < E < E_M^*$, and thus for the overall amount of charges loaded into the bulk between $\mu_1$ and $\mu_2$ [See Eq.~(6) in the main manuscript].}
	\label{fig:newmodel}
\end{figure}
\makeatother
In order to see this, consider a variation $\mu\to\mu+\delta\mu$.  From Eq.~\eqref{eq_toy_DOS} one obtains
the dimensionless TDOS
\ber
\frac{\partial \tilde{n}_0}{\partial \mu} &=& \frac{1}{\delta \mu} \left[\int_0^{\mu+\delta\mu}\,{\rm d}E\, -\int_0^\mu\,{\rm d}E\, \right] = 1,
\nn\\
\frac{\partial \tilde{n}_T}{\partial \mu} &=& \frac{1}{\delta \mu}
\left\{
\int_0^{\mu+\delta\mu}\,{\rm d}E\,\delta f_{\mu+\delta\mu} + \int_{\mu+\delta\mu}^\infty\,{\rm d}E\,
\left[
1 + a(\mu+\delta\mu)(E-\mu-\delta\mu)
\right]\delta f_{\mu+\delta\mu}
+
\right.
\nn\\
&&
\quad
\left.
-\int_0^\mu\,{\rm d}E\,\delta f_\mu + \int_\mu^\infty 
\left[
1 + a\mu(E-\mu)
\right]\delta f_\mu
\right\}
\nn\\
&=&
\frac{1}{\delta\mu} \left\{
\underbrace{\int_0^\infty\,{\rm d}E\, \delta f_{\mu+\delta\mu}}_{= 0} - 
\underbrace{\int_0^\infty\,{\rm d}E\, \delta f_\mu}_{= 0} +
\right.
\nn\\
&&
\quad
\left.
a(\mu+\delta\mu)\int_{\mu+\delta\mu}^\infty\,{\rm d}E\,[E-(\mu+\delta\mu)]\delta f_{\mu+\delta\mu} -
a\mu\int_\mu^\infty\,{\rm d}E\,[E-\mu]\delta f_\mu
\right\}. 
\eer

The integrand $[E-(\mu+\delta\mu]\delta f_{\mu+\delta\mu}$ is only a 
	function of $E-(\mu+\delta\mu)$, ergo the change of variable 
	$\mu+\delta\mu\to\tilde\mu$ leads to
	\be
	\int_{\mu+\delta\mu}^\infty\,{\rm d}E\,[E-(\mu+\delta\mu)]\delta 
	f_{\mu+\delta\mu} =
	\int_{\tilde\mu}^\infty\,{\rm d}E\,[E-\tilde{\mu}]\delta f_{\tilde{\mu}} =
	\int_\mu^\infty\,{\rm d}E\,[E-\mu]\delta f_\mu \equiv I(T),
	\ee

which yields
\be
\frac{\partial \tilde{n}_T}{\partial\mu} = a\,I(T).
\ee
We thus obtain
\be
\frac{\partial \tilde{n}}{\partial\mu} = 1 + a\,I(T).
\ee

Note that such a result does not qualitatively change if one generalizes Eq.~\eqref{eq_toy_DOS} to
\be
\label{eq_super_toy_DOS}
\tilde{D}'(E,\mu) = 
\left\{
\begin{array}{ll}
	\delta g(E-\mu) & E_m^* < E \leq \mu, \\
	\delta g(E-\mu)[1 + a\mu(E-\mu)] & \mu < E < E_M^*,
\end{array}
\right.
\ee
where $\delta g(E-\mu)$ is a function symmetric around $\mu$ and such that $\delta g(E-\mu) \to 1$
for $|E-\mu|\gg k_BT$, which represents the Coulomb gap \cite{EfrosShklovskii1984s,CoulombGapTheorys}.
Using Eq.~\eqref{eq_super_toy_DOS} would only change the value of the integral $I(T)$,
at the price of requiring further assumptions and introducing additional parameters.
While a quantitative theory would clearly require taking the Coulomb gap into account,
the simpler toy model defined by Eq.~\eqref{eq_toy_DOS} is already enough for our discussion
and we limit ourselves to it.

\section{Discussion on frequency dependence}

\label{fdependence}
One needs to measure at low frequencies in order to avoid resistive effects, meaning that the measured signal no longer reflects the pure capacitive signal and thus the (thermodynamic) density of states. As emphasized in the main text, we measured at the lowest frequency of 50 Hz in order to avoid such resistive effects\cite{KozlovPRL2016_QCs}. Fig. ~\ref{fig:fdependence} shows $C(V)$-data at 50 Hz, 240 Hz and 20 kHz. The overall dependence on frequency is weak. At the highest frequency of 20 kHz our capacitance bridge allows, the $C(V)$-trace is shifted a bit down but the shape is preserved. This means that the higher frequency has the same effect as a slightly reduced parasitic capacitance. Overall, resistive effects are not very pronounced in this material system. They are e.g., much weaker than at the charge neutrality point in the 3D-topological insulator HgTe\cite{KozlovPRL2016_QCs}.

\makeatletter 
\renewcommand{\thefigure}{S\@arabic\c@figure}
\begin{figure}
	\includegraphics[width=.6\linewidth]{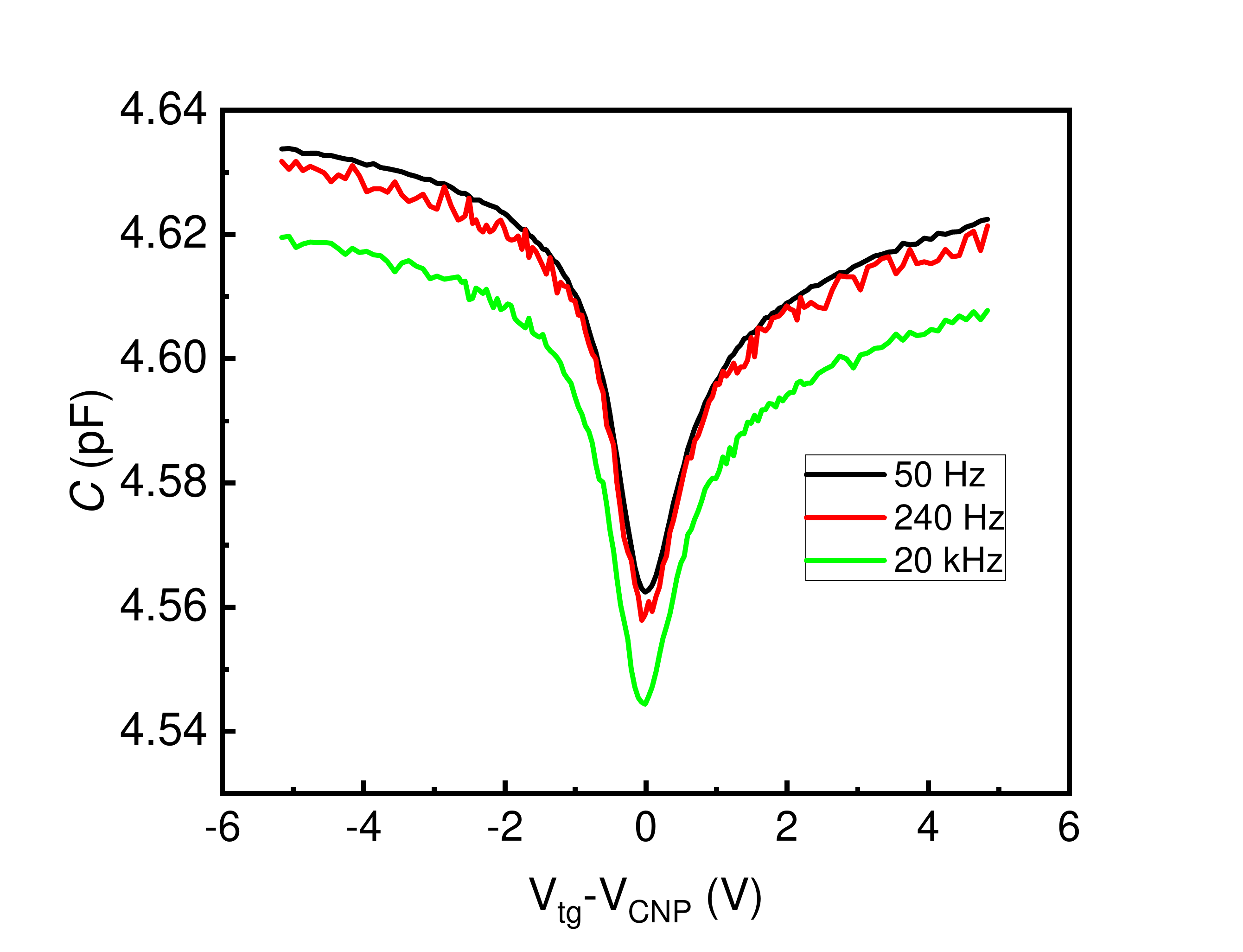}
	\caption{$C-V$ traces measured at three different frequencies 50 Hz, 240 Hz, and 20 kHz at 1.8 K, respectively. At the highest frequency, the trace is shifted towards lower capacitance values while maintaining the shape. }
	\label{fig:fdependence}
\end{figure}
\makeatother

\section{Discussion on hysteresis}

\label{C_hysteresis}
As mentioned in our main text we take sweeps in one direction only to avoid that hysteresis affects the measurements. Fig.~\ref{Figure_Chysteresis}a displays up-and down sweep of a $C(V_\text{tg})$ trace with clear hysteretic behavior, which we ascribe to charges in the insulator or at the insulator/BSTS interface. The traces are shifted back and forth depending on the sweep direction. In our paper we shift the voltage scale with the origin at the charge neutrality point (CNP). In Fig.~\ref{Figure_Chysteresis}b we show, by shifting up- and down-sweeps such that they coincide, that the traces are nearly on top of each other. Especially at small voltages between -0.5 and +0.5 V at which the Landau gaps occur, the traces agree well, thus justifying our approach to plot the data on a $V_\text{tg}-V_\text{CNP}$ scale. Hence, the hysteresis does not affect our analysis in the main text. Upon increasing $T$, the hysteresis stays nearly constant before it decreases at elevated $T$ and reaches at 90 K about 40\% of its low temperature voltage shift. This we checked in another, similar device.
We note that the $C(V)$ measurements in (Bi$_{1-x}$Sb$_x$)$_2$Te$_3$ (BST) \cite{WangPSSB2019s} are carried out in a very different regime. In the BST material investigated there, the Fermi-level is in the conduction band while in the present work it is in the bulk gap of BSTS. In BST the situation is similar to $n$-doped Si where the capacitance value depends on whether the system is under accumulation or inversion. The shape of the $C(V)$ traces are looking completely different compared to the one investigated here. In BSTS we have, in contrast to BST, transport which is dominated by the surface states at low temperatures. The  $C(V)$ traces look thus similar to the ones explored in graphene (apart from the asymmetry with respect to the CNP) or in HgTe \cite{KozlovPRL2016_QCs}.  

\makeatletter 
\renewcommand{\thefigure}{S\@arabic\c@figure}
\begin{figure}
	\includegraphics[width=0.9\linewidth]{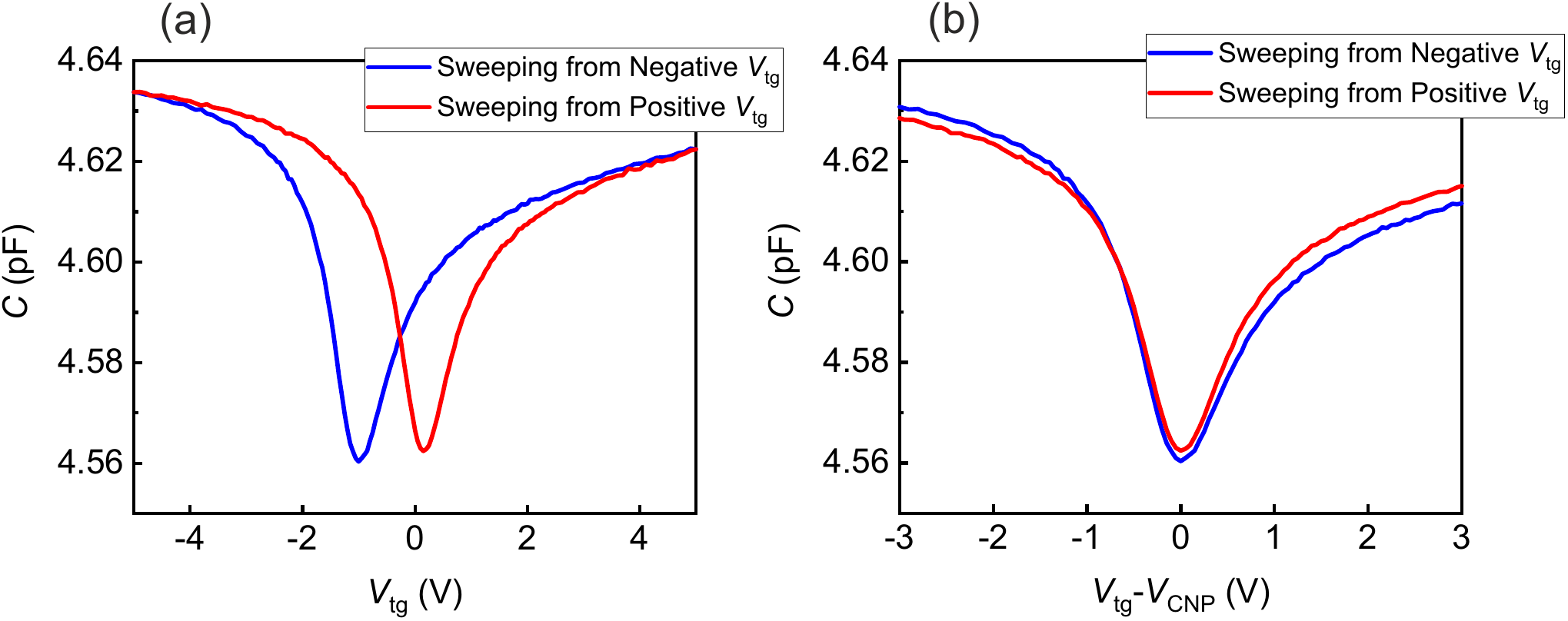}
	\caption{Detailed hysteresis analysis at 1.8 K and $V_\text{bg}$ = 0 V of the same device as in the main text. (a) $C-V_\text{tg}$ traces swept from both negative and positive $V_\text{tg}$ sides. (b) The up- and down-sweeps are shifted by $V_\text{CNP}$ such that the charge neutrality point is located at the zero of the voltage scale.} 
	\label{Figure_Chysteresis}
\end{figure}
\makeatother

%

\end{document}